\documentclass{article}
\usepackage[utf8]{inputenc}
\usepackage{csquotes}
\usepackage[english]{babel}
\usepackage[backend=biber,style=authoryear,sorting=nyt]{biblatex}
\usepackage{amsmath}
\usepackage{amssymb}
\usepackage{listings}
\usepackage[margin=1in]{geometry}
\usepackage{graphicx,float}
\graphicspath{ {./images/} }

\title{Bayesian Learning: A Selective Overview }
\author{
Yu Lin Hsu,\\ Chu Chuan Jeng,\\  Pavithra Sripathanallur Murali,\\ Mohammadreza Torkjazi,\\ Jonathan West,\\ Michaela Zuber,\\ Vadim Sokolov\thanks{All authors are of George Mason University. Corrisponing author's email: vsokolov@gmu.edu}}
\date{First Draft: Oct 27, 2021\\This Draft: Dec 22, 2021}

\begin{document}
\maketitle
\begin{abstract}
    This paper presents an overview of some of the concepts of Bayesian Learning. The number of scientific and industrial applications of Bayesian learning has been growing in size rapidly over the last few decades (\cite{damien2013bayesian}). This process has started with the wide use of Markov Chain Monte Carlo methods (\cite{geman1984stochastic,gelfand1990sampling,gamerman2006markov}) that emerged as a dominant computational technique for Bayesian in the early 1990's. Since then Bayesian learning has spread well across several fields from robotics and machine learning to medical applications. This paper provides an overview of some of the widely used concepts and shows several applications. This is a paper based on the series of seminars given by students of a PhD course on Bayesian Learning at George Mason University. The course was taught in the Fall of 2021. Thus, the topics covered in the paper reflect the topics students selected to study. 
\end{abstract}

\section{Introduction}
Bayesian learning combines probabilistic and statistical modeling and computational algorithms to draw conclusions from data. Given data $Y$ and a probabilistic model of the data $p(y\mid \theta)$ that is parameterized by $\theta$, in Bayesian learning we assume that the parameter we are to estimate is a random variable and we need to specify which probability distribution it follows. Thus, in Bayesian approach, besides likelihood we need to specify the probabilistic model of the parameters  $p(\theta)$, called the prior. Then, Bayesian estimate is calculated using the Bayes rule
\[
p(\theta \mid y) = p(y\mid \theta)p(\theta)/p(y).
\]

Here $\theta = (\theta_1,\ldots,\theta_p)$, and the total probability is 
\[
p(y) = \int  p(y\mid \theta)p(\theta) d\theta
\]
It is a normalizing constant that makes posterior a proper density function, e.g. to integrate to 1. Sometimes we will write 
\[
p(\theta \mid y) \propto p(y\mid \theta)p(\theta),
\]
posterior is proportional to the product of likelihood and prior. 

The prediction can be performed using
\[
p(y\mid D) = \int p(y,\theta\mid D)d\theta = \int p(y\mid \theta,D)p(\theta\mid D)d\theta
\]

Maximum likelihood estimation procedure is called frequentest, since it only accounts for the frequencies of the events observed in the data. While Bayesian approach combines that with prior assumptions about model parameters. The underlying difference between Bayesian and classical (a.k.a frequentist) approach to data analysis is the way uncertainty is treated. 

	To demonstrate the difference in the approaches, we estimate fairness of the coin based on the outcome of two flips. Say we have Head both times. The frequentest method will count number of heads $m = 2$ (thus the name of the method) and will divide by the number of experiments $n = 2$ to conclude that the coin is single-sided. However, this conclusion does not seem to coincide with what we know about coins. We have seen many coins before and all of them were fair two-sided ones. What are the chances that this one is different? In Bayesian approach we will have some prior assumptions, say that coin is fair and will update this assumptions based on the outcome of two experiments to obtain that there is small probability that coin is single-sided. We will show formal calculations for obtaining those updated probabilities, called posteriors and ways to encode prior assumptions via distribution functions called priors. However, when we flipped coin thousand times $m = 1000$ and got Head on every single flip, concluding that coin is single-sided now is logical. This also illustrate the asymptotic nature of the maximum likelihood estimator. 

Let's have a more rigorous analysis of the coin flipping example. Say we flipped coin $n$ dimes and observed a sequence of heads and tails denoted by $D$. We are interested in estimating the fairness of the coin $\theta$ (probability that head comes up). The likelihood for this experiment is 
\[
p(D\mid \theta) = \theta^k(1-\theta)^m
\]
where $k$ is the number of observed heads and $m$ is the number of tails, $n = k+n$. MLE estimate is $\theta = k/n$. To define a Bayesian model, we need to define a prior over $\theta$. We will use $Beta$ distribution which is defined on $(0,1)$ and is a generalization of a uniform distribution

The main difference when compared to classical approach is the subjective view of probability used in Bayesian analysis. The subjective means that different analyst might have different assumptions or different level of knowledge about unknown quantities used in a model. An objective uncertainty means that we have no assumptions about how uncertain a quantity is and can make conclusions regarding the uncertainty only based on the data. However, in real life an objective approach can be improved by introduction subjective knowledge, i.e. using Bayesian approach. Think of ten  coin flips. An objective approach does not allow us to say anything about the outcome of those 10 coin flips. We can only make conclusions about, say fairness of the coin, based on the outcome of the experiment On the other hand, subjectively we should ``suspect'' that we will see around five heads and five tails. Most of the quantities that we call random are such due to our lack of knowledge or inability to obtain required measurements. For example, an outcome of the coin flip is governed by the laws of Newtonian mechanics. If we are to know weight of the coin, its initial speed and number of rotations per second, then using simple motion equations, we can predict whether it is to land Head or Tails. Thus, the coin toss is a deterministic process. If we are to have access to measurements about the  speed and the rotation frequency we can make a much better guess about the outcome of a coin flip. The accuracy of our H/T guess will depend on how accurate our measurements about initial states are. From probabilistic point of view a certain quantity is just a degenerate case with probability density being a delta function
\[
\delta(x) = \left\{\begin{array}{cc}
1 & x=X \\ 
0 & x\ne X.
\end{array} \right.
\]

Whenever we have uncertainty, our density functions becomes non-degenerate. Thus, the density function allows us to encode our subjective uncertainty about a quantity. 

Now the question is, do we ever have objective uncertainty, or a random variable always a deterministic quantity inaccurately measured? Two examples include quantum mechanics, nuclear decay. Nuclear decay is impossible to predict and is impossible to predict according to our current physical knowledge of this process. In fact, the outcomes of nuclear decay process are used by on-line poker rooms as the most reliable source of random numbers. Some also can argue that human decision making is a random process. A relatively new interdisciplinary field called neuroeconomics studies the question of randomness of our decisions. 

In frequentist approach we distinguish deterministic and random variables. For example, when we want to calculate a mean given a sample. Then, the mean itself is deterministic, but estimate of the mean is random due to sampling errors. In Bayesian approach all of the quantities are random. The randomness is due to our lack of knowledge. To model that we will assume that the mean is random and can assign a density function to encode our uncertainty about this quantity. Thus we can use probabilistic approach to deal with quantities that appear to be deterministic. Even, when quantity is truly deterministic, we can use delta function as probability density. For example, in quantum mechanics, we have collapse of a wave function. The bottom line, we assume uncertainty about every quantity we deal with. 

Maximum likelihood has many good statistical properties and many theoretical results were obtained to show effectiveness of this approach in many situations. However, those results are asymptotic, i.e. assume that the sample size goes to infinity. In practice, it means we need to have large number of samples before those nice properties ``kick-in''.

\section{Hypothesis Testing}

Hypothesis testing and significance is the corner stone of any statistical analysis; it is what we use to determine the meaning of the results of our experiments. It is a guide to further research and study. And it is what we use to inform decision makers about courses of action to take based on observed data.

Hypothesis testing, as we know it today, is typically performed via Ronald Fischer's p-values and Egon Pearson's Type I and Type II errors. These methods, developed in the 1920's, made significance testing and data analysis straightforward and accessible in the scientific community. Here is an example of hypothesis testing with p-values. Suppose we have data of 100 coin flips, where 63 have come up heads. We want to determine whether or not the coin is fair, and our null hypothesis $H_0: p = 0.50$ is that the coin is fair  and alternative hypothesis is that the coin is not fair $H_1: p \neq 0.50$. Thus, under the null hypothesis, we assume that observations Bernoulli distribution with the mean of 0.5; $x\sim Bin(0.5)$. Now we compute the standard deviation, $\sigma$, of this distribution.
$$
\sigma = \sqrt{\frac{p(1-p)}{n}} = \sqrt{\frac{0.50\times 0.50}{100}} = 0.05.
$$
Now we take the estimated probability of heads from our data sample, $\hat{p} = \frac{63}{100}$ and compute the $z$-score.

$$
z = \frac{\hat p- \mu}{\sigma} = \frac{0.63 - 0.50}{0.05} = 2.6
$$
Due to central limit theorem, we have $z\sim N(0,1)$ and thus, $P_{\text{std norm}}(|z| > 2.6) = 0.00932$. If we assume a significance level of 0.05, then we reject the null hypothesis and conclude there is not enough evidence to suggest the coin is fair.

Though this methodology became widely used and accepted seemingly overnight, it was not absolved from scrutiny, specifically from Bayesian statisticians. The most prominent Bayesian at the time, Harold Jefferys, gave the criticism ``the p-value includes only data that did happen, not anything that might have happened but didn't`` (\cite{lindley_philosophy_2000}). To absolve this error, the Bayesians developed their version of hypothesis testing. This method was focused on finding the posterior probability of the null hypothesis itself, rather than the probability of extreme data occurring. Below is an example of a Bayesian hypothesis test with the same fair coin. 

The likelihood for this data is binomial
\[
P_{\text{x heads}} = {n\choose x} p^xq^{n-x},
\]
where $n$ is the total number of flips, $x$ is the number of heads, $p$ is the probability of heads on a single trial, and $q$ is $1-p$.

Lets denote this probability to be $P(heads) = \theta$ given what we observe in our data. So we want to find $P(\theta|D)$ where $D$ is our data of coin flips. Using Bayes theorem, we thus want to find
\[
P(\theta) = \dfrac{P(D\mid\theta)P(\theta)}{P(D)}.
\]

First we need to choose a prior distribution for $\theta$. Since we are modeling a binomial distribution, we can choose a Beta distribution to maintain conjugacy. To not bias the posterior, we choose the prior to be Beta($\alpha$ = 2, $\beta$ = 2).

\begin{figure}[H]
\centering
\includegraphics[scale = 0.4]{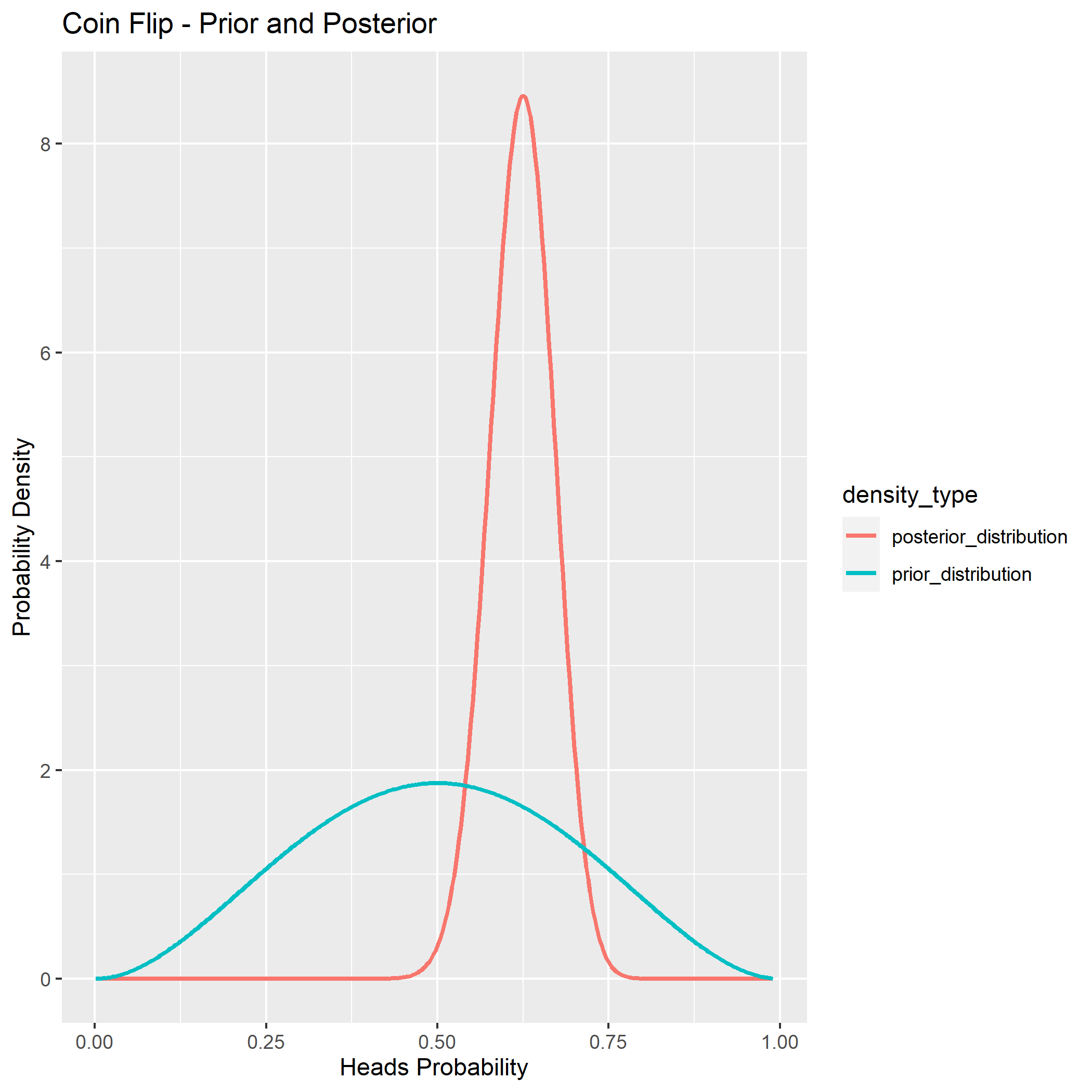}
\caption{Prior and Posterior Densities for the Coin Flip Example.}
\label{fig:betaflip}
\end{figure}

We update the parameters $\alpha$ and $\beta$ and obtain the posterior of Beta(65, 39), as shown in Figure \ref{fig:betaflip}.

We accept or reject the null hypothesis using an estimation approach or a comparison approach. The estimation approach uses a pre-defined Region of Practical Equivalence (ROPE) and compares to the posterior distributions high density interval (HDI). In our example of the fair coin, suppose we choose the ROPE to be a range of 0.45 to 0.55. That is, we will conclude the coin is fair if the ROPE is contained in the HDI of the parameter's posterior distribution. Calculating the HDI to be [0.53, 0.71], even though there is slight overlap, the null hypothesis is outside the lower bounds of the HDI and we find that we reject the null hypothesis and conclude that it is unlikely the coin is fair. 

The comparison approach uses the Bayes Factor to compare the null hypothesis to the alternative, finding which is more likely. The Bayes factor for the fair coin can be expressed as

$$
\frac{P(n,x|\theta_0 \neq 0.50)}{P(n,x|\theta_0 = 0.50)} = \frac{Beta(x + \alpha_{alt}, n - x + \beta_{alt})/Beta(\alpha_{alt}, \beta_{alt})}{\theta_0^x \times(1 - \theta_0)^{(n - x)}}.
$$

One caveat of using Bayes' factor is that the priors used for the alternative hypothesis have a large effect on the ratio. However, it is still common to use an uninformative prior. For this example, we will use the prior Beta(1,1). This gives us a ratio of 3.67, indicating that the alternative hypothesis is more likely to be true than the null. 

One would hope the results of the frequentist and Bayesian methods of hypothesis testing would converge - they use the same hypotheses and same data. Dennis Lindley, a fervent Bayesian in the mid 20th century, found that this is sometimes not the case.

\cite{lindley_statistical_1957} writes "...if $H$ is a simple hypothesis and $x$ the result of an experiment, the following two phenomena can occur simultaneously:

\begin{itemize}
  \item[(i)] a significance test for $H$ reveals that $x$ is significant at, say, the 5\% level;
  \item[(ii)] the posterior probability of $H$, given $x$, is, for quite small prior probabilities of $H$, as high as 95\%."
\end{itemize}

How can this happen? Even with the same data, the methods produce wildly different results. Lindley offers a more hypothetical example to show the main culprits of the dissonance. 

First let $x$ be a random variable drawn from a normal distribution with mean $\theta$ and variance $\sigma^2$. For simplicity, we will assume we know $\sigma^2$. Since we are using Bayesian analysis, we define the prior probability of $\theta$ to be some hypothesized value $\theta_0$ as $c$. Note, this $c$ is a probability, not a distribution. We define the rest of the distribution to be uniform over some interval $I$ which contains $\theta$. It may be helpful to think of this distribution as a flat line with a spike at $\theta_0$ - so $c$ of the distributions mass is concentrated at $\theta_0$. 

Now that we have our prior established, we can write the posterior probability that $\theta = \theta_0$ as

$$
\bar{c} = \frac{c\times\exp\{-n(\bar{x} - \theta_0)^2/(2\sigma^2)\}}{K}
$$

where

$$
K = c\times\exp\{-n(\bar{x} - \theta_0)^2/2\sigma^2\} + (1-c) \int_I \exp\{-n(\bar{x} - \theta)^2/2\sigma^2\}d\theta
$$

Since we want to show the resulting posterior is drastically different from the significance level, we need to choose an $x$ that is significant at the chosen significance level. We can express $x$ as

\[
\bar{x} = \theta_0 + \lambda_{\alpha}*\sigma/\sqrt{n}.
\]
Here, $\theta_0$ is our hypothesized value of $\theta$ and $\lambda_{\alpha}$ is a value dependent on $\alpha$ and found with the normal distribution tables. So we choose an $\bar{x}$ that is statistically different from $\theta_0$ at some significance level.

Plugging this value of $\bar{x}$ into our posterior probability equation, we find

\[
\bar{c} = \frac{c*exp\{-\frac{1}{2}\lambda_{\alpha}^2\}}{c*exp\{-\frac{1}{2}\lambda_{\alpha}^2\} + (1-c)\sigma * \sqrt{2\pi/n}\}}.
\]

Notice the second term in the denominator will trend to 0 as $n$ increases, forcing the entire expression to 1.

This is the root of the paradox. Our sample size $n$ matters. Lindley argues that the chosen significance level needs to depend on sample size. Much criticism also focuses on the use of a prior, and that the prior influences the discrepancy between the two methods. The choice of this prior, Lindley argues, is reasonable and intuitive. Often when an experimenter is designing their experiment, they have some reasoning to test whether or not the null hypothesis equals a certain value. In other words, the null hypothesis is special in some way. Moreover, mathematically, this prior, when chosen properly does not bias the posterior distribution.  The reason we use a uniform distribution over all other values in the interval $I$ is so that we don't bias the experiment to any other result. Normally, our value for $c$ is representative so that we also don't bias the experiment towards the null hypothesis as well. 

But what about other priors? James O. Berger and Thomas Selke (1987) discuss that there will always be a discrepancy between the p-values and posterior probabilities, no matter the prior used. 

Selke and Berger look into 4 families of prior distributions:

\begin{enumerate}
    \item 1. All distributions
    \item All distributions symmetric about $\theta_0$
    \item All unimodal distributions symmetric about $\theta_0$
    \item All $N(\theta_0, \tau^2)$ distributions, where $0 \leq \tau^2 < \infty$
\end{enumerate}

The authors examine the lower bounds of the posteriors of these distributions to determine if there is any overlap with p-values. The lower bounds can be expressed by

$$
\underline{Pr}(H_0 | x, G) = inf_{g \in G}Pr(H_0 | x)
$$

where $g$ is a specific distribution in the family of $G$ distributions. In this case, we are conditioning the null hypothesis on the data and the family of distributions the null hypothesis is being modeled with. Now we focus on the right side of this equation, specifically $Pr(H_0 | x)$. The intent is to generalize this enough so that different families of priors can be introduced and tested. First consider the marginal density of $x$. This can be expressed as

$$
m(x) = f(x|\theta_0)\pi_0 + (1-\pi_0)m_g(x)
$$
where

$$
m_g(x) = \int f(x|\theta)g(\theta)d\theta.
$$

We can express the posterior probability of $H_0$ as

$$
Pr(H_0 | x) = f(x|\theta_0) * \frac{\pi_0}{m(x)} = \left[ 1 + \frac{(1 - \pi_0)}{\pi_0} * \frac{m_g(x)}{f(x|\theta_0)} \right].
$$
Notice that $\frac{m_g(x)}{f(x|\theta_0)}$ is the Bayes factor for $H_0$ to an alternative $H_1$. If we denote this $\underline{B}(x,G)$ to take the lower bound, we are trying to find $\frac{f(x|\theta_0)}{sup_{g \in G} m_g(x)}$. Here, finding the least upper bound of $m_g(x)$ will give us the lower bound $\underline{B}(x,G)$. Including this representation into our posterior probability expression, we have

$$
\underline{Pr}(H_0 | x) =  \left[ 1 + \frac{(1 - \pi_0)}{\pi_0} * \frac{1}{\underline{B}(x,G)} \right],
$$
which gives us the lower bound of the posterior probability of $H_0$.

We can now algebraically deduce the impacts of different probability families on the posterior. The tables below show the results of these impacts.

\begin{table}[H]
\centering
\begin{tabular}{ |p{3cm}| p{3cm}|p{3cm}|  }
 \hline
 \multicolumn{3}{|c|}{All Distributions} \\
 \hline
 P Value ($p$) & $t$ & $Pr(H_0 | x, G_A)$\\
 \hline
 0.1   & 1.645 & 0.205\\
 0.05 & 1.96 & 0.128\\
 0.01 & 2.576 & 0.035\\
 0.001 & 3.291 & 0.0044\\
 \hline
\end{tabular}

\begin{tabular}{ |p{3cm}| p{3cm}|p{3cm}|  }
 \hline
 \multicolumn{3}{|c|}{All distributions symmetric about $\theta_0$} \\
 \hline
 P Value ($p$) & $t$ & $Pr(H_0 | x, G_A)$\\
 \hline
 0.1   & 1.645 & 0.34\\
 0.05 & 1.96 & 0.227\\
 0.01 & 2.576 & 0.068\\
 0.001 & 3.291 & 0.0088\\
 \hline
\end{tabular}

\begin{tabular}{ |p{3cm}| p{3cm}|p{3cm}|  }
 \hline
 \multicolumn{3}{|c|}{All unimodal distributions symmetric about $\theta_0$} \\
 \hline
 P Value ($p$) & $t$ & $Pr(H_0 | x, G_A)$\\
 \hline
 0.1   & 1.645 & 0.39\\
 0.05 & 1.96 & 0.29\\
 0.01 & 2.576 & 0.109\\
 0.001 & 3.291 & 0.018\\
 \hline
\end{tabular}

\begin{tabular}{ |p{3cm}| p{3cm}|p{3cm}|  }
 \hline
 \multicolumn{3}{|c|}{All $N(\theta_0, \tau^2)$ distributions} \\
 \hline
 P Value ($p$) & $t$ & $Pr(H_0 | x, G_A)$\\
 \hline
 0.1   & 1.645 & 0.412\\
 0.05 & 1.96 & 0.321\\
 0.01 & 2.576 & 0.133\\
 0.001 & 3.291 & 0.0235\\
 \hline
\end{tabular}
\end{table}

Notice that for each of the distribution families, the lower bounds are still much larger than the p-values for each t-statistic. Because these families of distributions comprise most priors used by experimenters, a discrepancy will very likely be observed. 

Why do the frequentist p-value approach and Bayesian posteriors give such different results, no matter the use of priors? One reason could be the assumptions each methodology makes before heading into the math. The philosophical differences between the two methods are outlined in Lindley's ``Tdche Philosophy of Statistics''.

Lindley's agrument can be summarized by the following:

\begin{itemize}
    \item[(a)] "Statistics is the study of uncertainty
   \item[(b)] Uncertainty should be measured by probability
   \item[(c)] Data uncertainty is so measured, conditional on the parameters
   \item[(d)] Parameter uncertainty is similarly measured by probability
   \item[(e)] Inference is performed within the probability calculus" (Lindley 2000)
\end{itemize}
   
Here we want to focus on argument (e), inference using probability calculus. The rules of probability calculus are

\begin{itemize}
    \item Convexity: for all $A$ and $B$, $0 \leq P(A|B) \leq 1$ and $P(A|A) = 1$.
    \item Addition: if $A$ and $B$ are exclusive, given $C$, $P(A \cup B|C) = P(A|C) + P(B|C)$
    \item Multiplication: for all $A$, $B$, and $C$, $P(AB|C) = P(A|BC)*P(B|C)$
    \item Conglomerability: if $\{B_n\}$ is a partition, possibly infinite, of $C$ and\\ $P(A|B_n C) = k$, the same value for all $n$, then $P(A|C) = k$.
\end{itemize}

Lindley argues that any inference that violates these rules is invalid and should not be used. In our case, we will focus on the significance levels and confidence intervals.

We can summarize a significance level to be the probability of the data from an experiment occurring, given that the hypothesis is true, $P(x|H)$, whereas the posterior probabilities calculate the probability of the hypothesis given the evidence of our data $P(H|x)$. Many scientists confuse the two, or think they mean the same thing. So despite the math that is used to calculate p-values and posteriors, the differing methodologies are actually searching for different things. So of course they are going to give different results. At this point, the problem becomes which probability should we be searching for to accept or reject our hypothesis? 

\cite{berger_could_2003} attempts to find reconciliation between the three major methods of hypothesis testing. In particular, he focuses this reconciliation to be a hypothesis test that addresses the major critiques of each school. For p-values, the major critique is that the p-values focus only on data that occurred and do not take into account more extreme data that did not occur in the experiment. The major critique of priors is that the use is too subjective and not objective enough. Most criticisms against the Bayesian methods are theoretical rather than empirical. And the major criticisms against Neyman-Pearson testing is that Type 1 and Type II errors do not reflect the variation in evidence as the data range over the rejection or acceptance region and also the need for alternative hypotheses.

Berger postulates a conditional test that would satisfy the three groups. For now, let us evaluate two hypotheses.

\begin{itemize}
    \item Condition on $S = \max\{p_0, p_1\}$
    \item reject $H_0$ when $p_0 \leq p_1$, accepting otherwise
    \item compute Type I and Type II conditional error probabilities using Bayes factor
    \item if $p_0 \leq p_1$, reject $H_0$ and report Type I CEP $\alpha(x) = \dfrac{B(x)}{1 + B(x)}$
    \item if $p_0 > p_1$, accept $H_0$ and report Type II CEP $\beta(x) = \dfrac{1}{1 + B(x)}$
\end{itemize}

In this test, p-values are used to measure the strength of evidence in the data. Since this is the value we are conditioning on, Berger argues Fisher would approve of this method. However, Fisher might still disapprove of the necessity of an alternative hypothesis. Bayesians would approve of the use of Bayes factors, and Neyman-Pearson users would be satisfied that this new test adheres to the frequentist principles. This new test also includes error probabilities with the data, addressing one of the major critiques of Type I/II testing.



\section{Learning and Predictions}
\subsection{Gibbs Sampler for Non-conjugate Models}
Although conjugate prior allows to compute posterior analytically, this assumption is not correct sometimes. \cite{carlin_inference_1991}  proposed Gibbs sampler for nonconjugate Bayesian models. 

Some prior error densities will affect the model error density to be non-normal, which will deviate from the normality assumption. The authors mention several prior error densities, such as t-family, double exponential, exponential power series, and logical error. However, Gibbs sampling can be calculated analytically in non-normal distributions and non-linear functions.

The authors select an error distribution by the Gibbs sampler. Assume $\mathcal{M}_i $ be an indicator of one type of error distribution, and the nonlinear model \[y_j=\mu_j+\epsilon_{ij},\] 
where 
\[\mu_j=\mathbf{f(x_j,\theta)}\boldsymbol{\beta}.\] 
We would have the log likelihood 
\[\log \ p(\mathbf{y}|\theta,\boldsymbol{\beta},\boldsymbol{\lambda},\sigma^2,\mathcal{M}_i)=-\frac{1}{2\sigma^2}(\mathbf{y}-\mathbf{F_\theta}\boldsymbol{\beta})^T\Sigma_i^{-1}(\mathbf{y}-\mathbf{F_\theta}\boldsymbol{\beta})-n\log\sigma-\frac{1}{2}\sum_{j=1}^n\log \lambda_j,\] 
where $\Sigma_i=\mathrm{diag}(\lambda_1,...,\lambda_n)$ and $\mathbf{F_\theta}=(\mathbf{f(x_j,\theta)})$. For the case, the authors can obtain the posterior distribution 
\[
p(\mathcal{M}_i|\theta,\boldsymbol{\beta},\boldsymbol{\lambda},\sigma^2,\mathbf{y})=\frac{p(\mathbf{y}|\theta,\boldsymbol{\beta},\boldsymbol{\lambda},\sigma^2,\mathcal{M}_i)p(\mathcal{M}_i|\theta,\boldsymbol{\beta},\boldsymbol{\lambda},\sigma^2)}{\sum_{k=1}^m p(\mathbf{y}|\theta,\boldsymbol{\beta},\boldsymbol{\lambda},\sigma^2,\mathcal{M}_k)p(\mathcal{M}_k|\theta,\boldsymbol{\beta},\boldsymbol{\lambda},\sigma^2)}.
\]
Hence, we can get the estimated of $p(\mathcal{M}_i|\mathbf{y})$ from the Gibbs sampler methods 
\[
\frac{1}{G} \sum_{g=1}^G(number \ of \ \mathcal{M}^{(g)}\ equal \ to \ i).\] 
We also can get the posterior mean of other parameters from the complete conditional distribution via Gibbs sampling.

To illustrate the selection method, the author implement it with a non-linear model of enzymatic reaction data. The response variable $y$ is the reaction rate of the enzyme, and the covariate is the Puromycin substrate concentration. The matrix model is \[\mu_j=\gamma+\frac{\alpha x_j}{\theta+x_j},\] 
where $\gamma, \alpha \in \mathbb{R}$, and $\theta \in \mathbb{R}^+$. They assume that there are two prior error densities. One is $t_2$ error, and the other is double exponential error. They have the complete conditional distribution for the model error densities. Implementing the Gibbs sampler 2500 times for each model. The result is the probability of $t_2$ error given the response $y$ is 0.73, while the probability of double exponential error given the response $y$ is 0.27. Hence, the $t_2$ error are more suitable. Besides, the marginal posterior density of $\alpha$, $\hat{p}(\alpha|\mathbf{y},\mathcal{M}_i)$ can be obtained. Without doing the selecting  an error distribution again, the authors also can get the marginal posterior density of the mean rate $\hat{p}(\delta|\mathbf{y},\mathcal{M}_i)$ when the Puromycin substrate concentration is 0.5 in ppm $\delta=E[Y|X=0.5]$.

After specifying the model and prior error density, Gibbs sampler can make non-conjugate Bayesian models easier to implement for the inference of the marginal posterior density and can help compare candidate error densities. 

\subsection{Bayesian Regularization}
\cite{polson_bayesian_2019} give an overview of the main concepts in model regularization and introduce different regularization models including Ridge, Lasso, Horseshoe, and their Bayesian equivalents.\\
To understand the regularization concept, two Machine Learning (ML) fundamental concepts, namely bias, and variance, should be learned first. Bias means how well the ML model can capture the true relationship between data points in the training set and variance means how minimum is the variability of predictions from the data points in the testing set. If the model is able to perfectly capture the relationship between data points in the training set, bias is almost zero. However, there will be an overfitting issue that results in high variability when the model is applied to the testing set. Thus, this is not a good ML model. A good model needs to have a reasonable amount of bias to approximately capture the true relationship between data points in the training set. Therefore, it will result in a low variance in the predictions. Regularization is used to find the best bias-variance trade-off. Regularization helps with:
\begin{enumerate}
    \item Preventing overfitting
    \item Lowering the variance between predictions and data points in the testing set
    \item Decreasing the sensitivity of the predicted values to small variations in the predictors by introducing an amount of bias to each predictor
\end{enumerate}

\subsubsection{Traditional regularization approach}
The traditional regularization estimator for a linear model is to minimize the following optimization problem:
\begin{align*}
    \min_{\beta} \sum_i f(y_i,\beta^Tx_i) \\
    \textrm{s.t.} \sum_j \phi(\beta_j)\le s
\end{align*}

Where $y$ is the model output, $x$ is an input, $\beta$s are model parameters (slopes in linear prediction model), $s$ is a hyper-parameter controlling the bias-variance trade-off, and the $\phi$ function assigns a regularization penalty to each model parameter. 
\cite{tihonov1963solution} proposed the Lagrangian form of the regularization equation 
\begin{equation}
    \min ||y-X\beta||_p^p + \lambda ||(\beta-\beta^0)||_q^q
\end{equation}
where $\lambda$ is the weight on the regularization penalty. Ridge regression ($p=q=2$) and LASSO ($p=2$, $q=1$) are two particular cases. The $L_0$ problem ($q=0$) leads to an NP hard optimization problem.

\subsubsection{Bayesian regularization approach}
The Bayesian regularization defines prior distribution for model parameters. This helps with capturing the uncertainty in parameter estimations. To obtain the best values for model parameters, the negative log of the posterior needs to be calculated by considering the log-likelihood and log of the prior for the model parameter.

\begin{equation}
    -\log p(\beta|X,y)=(1/2) \sigma_\epsilon^2 \sum_i (y_i - x_i^T\beta)^2 + \log p(\beta|\tau)
\end{equation}
Thus, regularization penalty is the log of prior distribution.
\begin{equation}
    \phi_\tau(\beta) \propto \log p(\beta|\tau)
\end{equation}

One typical approach in Bayesian regularization is to use normal scale mixture priors 
\begin{equation}
  \beta_i|\lambda_i,\tau_i \sim N(0,\tau_i^2\lambda_i^2),
\end{equation}

If the Normal distribution is considered for the prior on $\beta$, the final regularization method will be Ridge regularization. If the Laplace prior is considered, the regularization will be of type Lasso, and if half-Cuachy prior is considered for $\beta$, the regularization will be of type Horseshoe. In the case of the horseshoe prior (\cite{carvalho2010horseshoe} )parameter $\tau_i$ shrinks model parameters to zero and the prior parameter, $\lambda_i$ prevents the shrinkage. Gibbs sampling can be quite helpful in applying Bayesian regularization.

To estimate the joint distribution of $\beta$, (\cite{carlin_inference_1991,carlin1992monte}) used Laplace prior and developed a Gibbs sampler for LASSO regularization.

In the Horseshoe regularization, $\tau$ is a global hyper-parameter and does not depend on index $i$, hence
\begin{equation}
    \beta_i|\lambda_i,\tau \sim N(0,\tau^2\lambda_i^2.
\end{equation}

Horseshoe is a global-local shrinkage prior which employs half-Cauchy distribution over the global and local parameters. There have been some problems with sampling from this prior which was addressed by multiple studies. An alternative approach approach is to use half-t priors instead of Cauchy when using small degrees of freedom.
\\

We compare different methods of regularization using a simulated data set $X \in R^{100*10}$. The data set is composed of 100 normally distributed samples for 10 regressors with mean zero and standard deviation of one. The true beta values were set to $\beta = (0.85,1.25,1.75,2.5,0,0,0,0,0,0)$. The output was also set to $y=X\beta + \epsilon$ with $\epsilon_i \sim N(0, 1)$. Such comparisons help with investigating differences between traditional methods and Bayesian regularization methods, and choosing the one that obtains more precise model parameters. \\

\begin{figure}[H]
    \centering
    \includegraphics[scale=0.75]{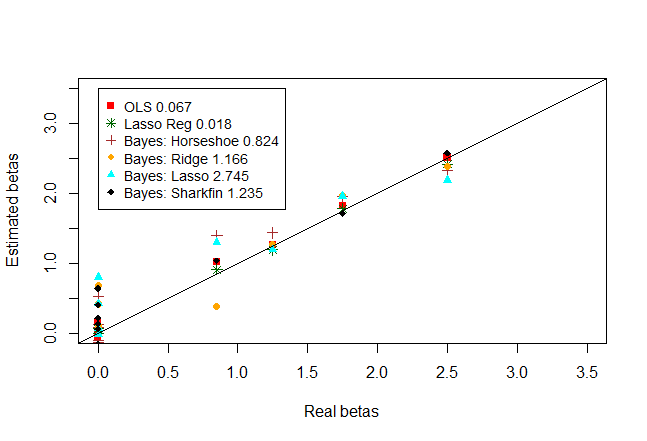}
    \caption{Betas estimated using OLS, regular Lasso, and posterior estimated betas using horseshoe, ridge, laplace, and sharkfin priors }
    \label{fig:my_label}
\end{figure}

The methods used in this comparison are Ordinary Least Square (OLS), Lasso Regression, and Bayesian regularization with Horseshoe, Ridge, Laplace, and Sharkfin priors. R language was chosen for this comparison. The linear regression model was used for the OLS and the ``glmnet'' package was employed for the traditional Lasso method. To gain the best result from the traditional Lasso, cross-validation was applied to investigate the best lambda value that is the minimum value for the penalty coefficient. To apply Bayesian regularization, the ``bayeslm'' package was used that supports the required priors for this comparison. To compare different methods, estimated betas with each method were compared against the true beta values and the squared errors were calculated and shown as numbers in the legend of Fig.1 to illustrate the performance of each method. The differences are shown in Figure \ref{fig:my_label}.

As can be seen, the traditional Lasso regression, which was applied with cross-validation, had the best performance with the lowest squared error value of 0.018 and Bayesian regularization with Laplace prior, which results in Bayesian Lasso, had the worst performance with the highest squared error value of 2.745. Among Bayesian approaches, Horseshoe prior demonstrated the best performance with the lowest error value of 0.824, and Laplace prior showed the worst performance with the highest error value of 2.745. The comparison also showed that the traditional Lasso regression outperforms the simple linear model by showing a better estimation of beta values and a smaller error value.

The main difference between the traditional Lasso approach and other methods, especially the Bayesian methods, is that the traditional Lasso regularization could perfectly eliminate 5 out of 6 betas with zero values. While other methods were unable to remove them. Even the Bayesian method with horseshoe prior, which had the best performance among Bayesian approaches, assigned a very small number to the zero-valued betas but did not eliminate them. Although more tests should be done before making any conclusion about the overall performance of each regularization method, however, the traditional Lasso outperformed all other regularization methods in our example and estimated beta values with the minimum error.

\subsubsection{Bayesian Assessment in Factor Analysis} 
The factor analysis is one of the powerful and flexible methods used to assess multivariate dependence and codependence at the stage of data understanding from the cross-industry process for data mining (\cite{lopes_bayesian_2004}). The cross-industry process for data mining follows the order of business understanding, data understanding, data preparation, modeling, evaluation, and deployment, respectively. After the stage of the business understanding, factor analysis could help understand the data by exploring the nature of the business, where various kinds of data relationships amongst the variables at study can be interactively verified or refuted. Due to the increased access to appropriate computational tools, such as iterative MCMC simulation methods, Bayesian factor analytic models were developed and applied in dynamic factor components with larger data sets in higher dimensions. The number of factors is essential for the modeling choice to explore the Bayesian inference in latent factor models. 

To explore MCMC methods for factor models under the number of the factors as unknown, (\cite{lopes_bayesian_2004}) developed a customized reversible jump Markov chain Monte Carlo algorithm (RJMCMC) for moving between models with different numbers of factors.Suppose we have the k-factor models, and the posterior distribution is
\[
p(\theta_k\mid y,k) \propto p(y\mid\theta_k,k)p(\theta_k\mid k)
\]
where $p(y\mid\theta_k,k)$ and $p(\theta_k\mid k)$ denotes the probability model and the prior distribution of the parameters of k-factor model. Then
\[
p(\theta_k,l\mid y) \propto p(k)p(\theta_k\mid k,y)
\] 
The RJMCMC methods involve Metroplis-Hastings type algorithms that moves a simulation analysis between models defined by $(k, \theta_k)$ to $(k', \theta_{k})$ with different dimensions $k$ and $k'$, where $k$ is the number of the factors. The resulting Markov chain simulations jump between such distinct models, and the algorithms are designed to be reversible so as to maintain detailed balance of the chain. Further, the marginal prior probabilities $p(k)$ can be specified over $k\in K$ 
, and then the RJMCMC analysis processes are as follows.

Firstly, choose a starting value of $k$. Set current values of $\theta_k$ to draw from the posterior $p(\theta_k\mid k,y)$ by using one (or more) steps of the MCMC algorithm based on past sampled values from this $k$-factor model. Note that this step produces both new sampled value of $\theta_k$ and the factors $F_k$, though only the former are used in exploring moves to models with other $k$ values.

Secondly, between model move step:
\begin{enumerate}
\item Draw a candidate value of the number of factor $k'$ from a proposal distribution defined by pre-specified transition probabilities $Pr(k'\mid k) = J(k \rightarrow k')$
\item Given $k'$, draw the parameters $\theta_k'$ from the distribution $q_k'(\theta_k')$ of equation,
        \[ 
q_k(\theta_k) \equiv q_k(\beta_k, \Sigma) = q_k(\beta_k)\prod_{i=1}^m q_k(\sigma_i^2),  k \in K 
\]
\item Compute the accept/reject ratio
\[ 
\alpha = min \Big\{ 1, \frac{p(y\mid k', \theta_{k'})p(\theta_{k'}\mid k')p(k')}{p(y\mid k, \theta_k)p(\theta_k\mid k)p(k)} \frac{q_k(\theta_k)J(k'\rightarrow k)}{q_k'(\theta_{k'})J(k\rightarrow k')} \Big\}
\]
For each value $j \in (k, k')$, $p(y\mid j,\theta_j)= p(y\mid j,\beta_j,\Sigma)$ is the likelihood function in equation as below: 
\[ 
p(y\mid\beta,\Sigma) \propto |\Omega|^{-T/2}etr(-0.5\Omega^{-1}y'y)
\]
,$p(\theta_j\mid j)$ is the prior density function for the parameters within the j-factor model, and $p(j)$ is the prior probability on $k$ factors. With probability $alpha$, accept the jump to the $k'$-factor model and the new parameter values $theta_k'$ just sampled as candidates. 
\end{enumerate} 

Thirdly, if the jump to model $k'$ accepted, run one step of the MCMC analysis in this $k'$-factor model, producing new sample values of the full set of quantities $(\theta_{k'}, F_k')$. Otherwise, remain in model $k$ and use the MCMC to produce new values of $(\theta_k, F_k)$. Lastly, repeat 1. and 2. until practical convergence is judged to have been achieved.


\subsection{Bayesian Interpolation}
In this study, \cite{mackay_bayesian_1992} introduces levels of inferences and the Bayesian approach for models comparison to choose the best ML model that describes the data set better than other models. The author demonstrates the Bayesian model comparison by applying it to an interpolating of noisy dependent variables. It is proved that the model comparison can be done by comparing the likelihoods of different models where the priors are the same for all models.

It is common to decompose inference, such as interpolating, learning and classification, into two levels
\begin{enumerate}
    \item Inferring that what values the model parameters should take given data
    \item Comparing and ranking different fitted models to choose the best one given data
\end{enumerate}

The results of the Bayesian method for the first inference are almost close to the ones from the frequentist approaches. But for the second inference, model comparison, there is no frequentist method to solve it. It is not the task of choosing the model that fits the data best, because it may lead us to over-parametrized and complex models. The main idea of this paper is to find the model that better does the second inference task for the interpolation example of a given data set.

The Bayesian approach for model comparison is using the Occam's razor principle: ``It is not necessary to use complex models when there are simpler models to fit the data''. The Bayesian approach employs this principle and makes the complex models to self-penalize themselves which makes their rank worse in comparison with simpler models. 

\subsubsection{Applying Bayes rule for two inference levels}
The main task of model fitting is identifying the best value for model parameters $w$ 
\begin{equation}
    P(w|D,H_i)=\frac{P(D|w,H_i)P(w|H_i)}{P(D|H_i}
\end{equation}

where $H_i$ is model $i$ and D is the data set.
The main task of model comparison is investigating the best model among fitted models using the following Bayes rule 
\begin{equation}
    P(H_i|D) \propto P(D|H_i)P(H_i)
\end{equation}

Considering a similar prior for all models, the only term that distinguishes models is the likelihood (evidence) of each model. Thus, evidences are used to evaluate models and define their ranks. The Bayes rule is not normalized because new models may be developed once new data arrives. So, it's open-ended. New models can be compared to the old ones according to the new data.

The evidence of the above equation is the normalizing constant of the model-fitting equation. 
\begin{equation}
    P(D|H_i)=\int P(D|w,H_i)P(w|H_i)dw \simeq P(D|w_{MP},H_i)P(w_{MP}|H_i)\bigtriangleup w
\end{equation}

where $w_{MP}$ is the model parameter that results in the maximum posterior distribution. It is the peak of probable parameter $w$ that can be used to approximate the evidence by multiplying its height by the width of parameter $\bigtriangleup w$. The $P(w_{MP}|H_i)\bigtriangleup w$ is the Occam factor part for one parameter $w$. Occam factor penalizes H models for having parameter $w$. Where there are multiple parameters with wide ranges for a complex ML model, the Bayes rule results in a larger Occam factor, hence, more penalty for the complex model. Thus, the Occam factor measures the model's complexity, and the prior term acts as a regularizer when comparing models.

\subsubsection{Interpolating: First level of inference}
Likelihood function:\\
\begin{equation}
    P(D|w,\beta,A)=\frac{exp[-\beta E_D(D|w,A)]}{Z_D(\beta)}
\end{equation}
Prior function:\\
\begin{equation}
    P(w|\alpha,A,R)=\frac{exp[-\alpha E_w(w|A,R)]}{Z_w(\alpha)}
\end{equation}
where $E_w$ is the regularizing function and $\alpha$ is regularizing constant. \\
Thus, the posterior is:
\begin{equation}
    P(w|D,\alpha,\beta,A,R)=\frac{p(D|w,\beta,A)p(w|\alpha,A,R)}{p(D|\alpha,\beta,A,R)}
\end{equation}

To simplify, the author considers $M(w)=\alpha E_w+\beta E_D$ as a new objective. Minimizing the combined $M$ function gives the most probable interpolant $w_{MP}$. The posterior then will be \\
\begin{equation}
    P(w|D,\alpha,\beta,A,R)=\frac{exp[-M(w)]}{Z_M(\alpha,\beta)}
\end{equation}

The next step is to identify the values for $\alpha$ and $\beta$. $\alpha$ is the regularizing constant coming from the prior and $\beta$ is the data parameter coming from the data set. $\alpha$ acts the same as the regularizing constant. A very large alpha value results in a smooth and flat interpolant that cannot capture all the data points and has a lot of bias. As $\alpha$ value decreases, the interpolant tries to fit the data better and a very small alpha value results in wild oscillating of the interpolant and over-fitting the noise in the data. Therefore, a good model needs a reasonable value of $\alpha$. To simplify the process, the author assumed a constant $\beta$ value. The best source to explore for  $\alpha$ and $\beta$ values is the data set. Hence,
\begin{equation}
    P(\alpha,\beta|D,A,R)=\frac{p(D|\alpha,\beta,A,R)p(\alpha,\beta)}{p(D|A,R)}
\end{equation}
As can be seen, the normalizing constant of two previous equation, $p(D|\alpha,\beta,A,R)$, became the evidence of the second level inference calculations. In addition to this, a flat prior was assumed for the values of $\alpha$ and $\beta$ to demonstrate that there is no information about them. Hence, the priors are the same for all models and what matters here is the evidence (likelihood) that can be represented in terms of normalizing constants as following 
\begin{equation}
    P(D|\alpha,\beta,A,R)=\frac{Z_M(\alpha,\beta)}{Z_w(\alpha),Z_D(\beta)}
\end{equation}

The evidence can also be represented by
\begin{equation}
    P(D |A,R) \simeq P(D |A,R,\hat{\alpha},\hat{\beta}) P(\hat{\alpha},\hat{\beta})2\pi\bigtriangleup\log\alpha\bigtriangleup\log\beta
\end{equation}

The authors demonstrated the potential of the Bayesian model comparison approach by comparing its application on two different data sets using three different models: Legendre polynomials, Radial basis, and Splines. The Occam factor worked well for the Legendre polynomials and its results chart shows a slight decrease in the evidence value after a specific number of model parameters. This proves that the Occam factor tries to decrease the rank of complex models that have multiple parameters. The same happened in the other two models. For example, the Radial basis model did not improve after including 50 parameters and the Splines model demonstrated its best performance with 3 coefficients. Calculated maximum evidence values for models helped with ranking models after being applied to these two data sets. 
Therefore, using evidences coming from the data, models can be ranked and regularizing constants, $\alpha$s, are found according to the maximum evidence values. The regularization is embedded in the Bayesian approach of model comparison and can act automatically to penalize the complex models. The author believes that the Bayesian method reduces the computational cost in comparison with the cross-validation method.

\subsection{Deep Learning: A Bayesian Perspective}

\cite{polson_deep_2017} review Deep Learning (DL) algorithm, its concept,  applications, and difference from other ML models, and the Bayesian Deep Learning algorithm. Using Bayesian methods in DL introduces new research opportunities to explore faster stochastic algorithms, tuning hyper-parameters, constructing good predictors, and interpreting models. 

Deep Learning is a form of machine learning that uses hierarchical layers to be used for high dimensional data sets to perform prediction and it can be successfully used for 
\begin{enumerate}
    \item Image processing
    \item Learning in games
    \item Neuroscience
    \item Energy conservation
    \item Skin cancer diagnostics
\end{enumerate}

\subsubsection{Deep Learning Fundamentals}
DL models create multiple layers to better refine the inputs and have more precise prediction or classification. To accomplish this goal, DL models use $F(X)$ as a multivariate function which is the superposition of multiple univariate functions $f_l$. Univariate activation functions are used to decompose the high-dimensional data set. Each layer has its own univariate function, hence,
\begin{equation}
    f_l^{w,b}=f_l(\sum_1^{N_j} W_{ij}z_j+b_l),
\end{equation}
where $W$ is the weight, $z$ is the input to layer $l$, and $b$ is the bias term. The DL predictor for $L$ layers is 
\begin{equation}
    \hat{Y}(x):=(f_1^{w_1,b_1} \circ ... \circ f_l^{w_L,b_L})(X)
\end{equation}

The challenge here is to identify Z variables and estimate the f functions correctly.

DL follows a hierarchy approach in which the output of each level will be used as input to the next layer.

\begin{gather*}
    Z^{(1)} = f^{(1)}(W^{(0)}X+b^{(0)}) \\
    Z^{(1)} = f^{(2)}(W^{(1)}Z^{(1)}+b^{(1)}) \\
    \ldots \\
    Z^{(L)} = f^{(L)}(W^{(L-1)}Z^{(L-1)}+b^{(L-1)}) \\
    \hat{Y}(x) = W^{(L)}Z^{(L)}+b^{(L)}
\end{gather*}

where $X=Z^{(0)}$ and the choice of univariate functions determines the performance of a predictor. The most commonly used deep learning architectures are feed forward, auto-encoder, convolution, recurrent, long/short term memory, and neural turing machines.

Traditional statistical techniques that deal with high-dimensional datasets are 
\begin{enumerate}
    \item PCA (Principal Component Analysis)
    \item PLS (Partial Least Squares)
    \item LDA (Linear Discriminant Analysis)
    \item RRR (Reduced Rank Regression)
    \item PPR (Project Pursuit Regression)
    \item Logistic Regression
\end{enumerate}

Traditional techniques usually use only two layers for decomposition and reducing the dimension, hence, they are considered as shallow learners in comparison with DL that includes deeper layers. The hierarchy in DL with multiple layers helps in refining a dataset more and reducing the input dimensions to the most important predictors.
\subsubsection{Bayesian DL}
In a Bayesian DL model, the output of a model can be obtained by optimizing the following 

\begin{equation}
    L_\lambda(Y,\hat{Y})=-\log p(Y|Y^{\hat{W},\hat{b}}(X)) - \log p(\phi(W,b)|\lambda)
\end{equation}
The first part is the negative log-likelihood and the second term is the negative log-prior distribution over parameters $W$ and $b$
\begin{equation}
    - \log p(\phi(W,b)|\lambda)=\lambda\phi(W,b)
\end{equation}

Bayes conditional averaging can be used for high-dimensional regression/classification. It constructs regions to reduce the dimensions. Using the Bayes method, DL partitions the input space like the one in decision trees, random forests, BART, CART, and MARS. For instance, 3 neurons in one layer of a DL model can divide the input space into 7 regions. The number of hyper-planes grows exponentially with the number of layers.



\subsection{BART: BAYESIAN ADDITIVE REGRESSION TREES}

The BART model and its fundamental basics were introduced in \cite{chipman_bart_2010}, this study and its performance was compared with other existing Sum-of-trees models to demonstrate its advantages. Sum-of-trees models are the ML models using the summation of several regression trees $g_j(x)$ for predictions or classifications. 
\begin{equation}
    Y=\sum_1^m g_j(x) + \varepsilon,\quad \varepsilon\sim N(0,\sigma^2)
\end{equation}

Existing linear models are
\begin{itemize}
    \item Boosting (fits a sequence of trees in which each tree is different by earlier ones)
    \item Bagging (randomly generates large number of independent trees and then averages the prediction)
    \item Random forest (randomly generates large number of independent trees and then averages the prediction)
    \item Bayesian model averaging (uses posterior probabilities of single-tree models as weights for averaging)
\end{itemize}

BART employs a prior that regularizes the fit and keeps the effect of each tree small. The difference between BART and the existing Bayesian approach is that in BART, the effects of each tree is weakened in a way that each tree explains a small and different portion of the additive function. BART also uses backfitting MCMC for sampling from the posterior of the model and making inferences. Thus, BART is a combination of a sum-of-trees model and a regularization prior.
The sum-of-trees model using m trees
\begin{equation}
    Y=\sum_1^m g(x;T_j,M_j) + \varepsilon,\quad \varepsilon \sim N(0,\sigma^2)
\end{equation}

where $g(x;T_j,M_j)$ is the function that assigns a $\mu_{ij} \in M_j$ for each tree $T_j$. Each $\mu_{ij}$ is a part $E(Y|x)$ and large number of trees improves BART's predictive capabilities. A regularizing prior should be defined for all model parameters $(T_j,M_j)$ and $\sigma$. The prior regularizes the model fit by lowering the influence of each tree on the final prediction. To have a prior that does not conflict significantly with the data at hand, the authors override this Bayesian rule of obtaining priors from previous experiences. The prior distribution is
\begin{gather*}
    p((T_1,M_1),...,(T_m,M_m),\sigma)=[\Pi_j p(T_j,M_j)]p(\sigma) \\
    = [\Pi_j p(M_j|T_j)p(T_j)]p(\sigma) \\
    = [\Pi_j p(\mu_{ij}|T_j)p(T_j)]p(\sigma)
\end{gather*}

\subsubsection{Regularization prior}
The prior has three different parts that will be discussed in the following.
$p(T_j)$ is specified by three aspects (according to \cite{chipman_bart_2010}):
\begin{itemize}
    \item The probability that a node at depth $d(=0,1,2,...)$ is non-terminal and can be calculated using $\alpha (1+d)^{-\beta}, \quad \alpha \in (0,1), \beta \in [0, \infty)$
    \item The distribution on the splitting variable at each interior node (uniform prior on available variables)
    \item The distribution on the splitting rule in each interior node (uniform prior on available splitting values)

\end{itemize}

$p(\mu_{ij}|T_j)$: To obtain this prior term:
\begin{itemize}
    \item The conjugate normal distribution $N(\mu_\mu,\sigma_\beta^2)$ will be used
    \item The $y$ values need to be transformed, so that, $y_{\min}=-0.5$, $y_{\max}=0.5$, $\mu_{ij} \sim N(0,\sigma_\mu^2)$, and $\sigma_\mu=\frac{0.5}{k\sqrt{m}}$
\end{itemize}

This prior shrinks the tree parameters $\mu_{ij}$ toward zero, hence, limits the influence of each tree components. As $m$ (number of trees) increases, the prior applies more shrinkage to the $\mu_{ij}$ parameters. The value for $k$ can be chosen using cross validation. However, the authors found the best value of $k$ to be between 1 and 3 and recommended $k=2$.
$p(\sigma)$: To obtain this prior term the conjugate prior of inverse chi-square distribution can be used. The authors recommend using a chi-square distribution with 3 degrees of freedom and 0.9 quantile. To calculate $m$ , we use the following algorithm
\begin{itemize}
    \item cross validation (computationally expensive)
    \item putting a prior on m and treat it as an unknown parameter (computationally expensive)
    \item Considering a default value and then checking whether one or two other values results in significant difference or not (the authors did it by considering $m=200$ as default)
\end{itemize}
As $m$ increases, the predictive performance of BART increases until some $m$ value that the performance starts to decrease gradually. Hence, $m$ should never be very small. A small $m$ makes the sum-of-trees model tries to include the $x$ components that are strongly related to $y$. This makes a competition between $x$ components for entry into the model.\\

\subsubsection{Making inferences from the BART posterior}
BART uses backfitting MCMC algorithm to sample from the posterior. It is a Gibbs sampler. The draw of $\sigma$ is from an inverse gamma distribution. Metropolis-Hastings can be used for the draw of $T_j$. And the draw of $M_j$ is a set of independent draws of the $\mu_{ij}$ of the terminal node from a normal distribution. The draws finally converge to the posterior distribution of the sum-of-trees model. The final prediction value will be the average or the median of all trees' predictions. The posterior is
\begin{equation}
    p((T_1,M_1),...,(T_m,M_m),\sigma | y)
\end{equation}

BART also can be used for making classifications where $Y=0 or 1$. In this case, the only difference would be the way the $p(\mu_{ij}|T_j)$ is obtained. For classification, the interval will become $(\Phi[-3],\Phi[+3])$, $\mu_{ij} \sim N(0,\sigma_\mu^2)$, and $\sigma_\mu=\frac{3}{k\sqrt{m}}$. \\

To demonstrate the potential of BART, the authors tested it performance in three scenarios: comparing results of application on 42 data sets with other ML models, generating the friedman's function and testing the BART's performance with different number of trees and predictors, comparing the accuracy of BART with other models in making classifications.

For the first scenario, the authors used a BART model using default values for parameters as mentioned before, and a BART model whose parameters are defined using cross-validation and compared their relative RMSE with Boosting, Neural Net, and Random Forest. The BART CV outperformed all models in all 42 data sets.\\

In the second scenario, the authors simulated a data set with 10 predictors and 100 observations and tried to investigate the sensitivity of BART to the number of trees and predictors. The Friedman's equation for the output is composed of 5 predictors only. So, the authors tried to test whether the BART can find the strongest predictors. \\

According to the test results, small number of trees resulted in poor RMSE and increasing number of trees beyond the required number results in a minor degradation of performance. To test the BART's potential in capturing true strongest predictors, they tested different values for predictors by simulating data. The in-sample estimates and their intervals were perfectly good for all values. However, for the out-of-sample, as number of predictors increases, the predictions shrinks toward the mean more, because there is less information about the data. 
The out-of-sample performance was also compared with other models (random forest, neural nets, and gradient boosting) by setting different number of predictors. Again, BART-CV outperformed all other models while BART with default values degraded significantly for 1000 predictors.\\

For the Friedman's test, BART execution time was also compared with other models by varying number of observations and number of predictors. The authors found that the number of samples influences the execution time exponentially while, the number of predictors is independent of the execution time. Considering 50 predictors, BART with default parameters values had the worst execution time. \\

In the third scenario, the accuracy of BART tested in classifying active compounds of cancer medications. The accuracy results of BART model were compared with other results from random forests, neural nets, and boosting methods. Random forests provided the best performance, then boosting, then BART, and finally support vector machines. \\

In summary, BART has three necessary parts:
\begin{itemize}
    \item sum-of-trees model
    \item Regularization prior
    \item Backfitting MCMC algorithm
\end{itemize}
Because of the prior regularization, each tree in the BART approach corresponds to only a part of the overall fit. BART CV outperforms other models in all cases. However, BART default is faster and easier and can be helpful in cases where there is a need for a less computationally expensive method.\\

\section{Surrogates and Optimisation}
A typical simulation modeling framework is composed of \textit{model}, \textit{calibration algorithm}, and \textit{analysis algorithms}. Simulators became a dominant model type used in many area, such as decease modeling (\cite{ozik2021population}) and public policy (\cite{kaligotla2018modeling,axtell2019frictional}) to transportation (\cite{sokolov2012flexible,auld2016polaris}). But a model is no good without a calibration algorithm to determine how to set its parameters and is very limiting without an optimization algorithm that allows to find an optimal configuration of the system design. The very same flexibility that makes simulation models appealing, also comes hand in hand with optimization problems that are intractable, with the number of simulations required to find an optimal solution growing exponentially, as the input dimension increases (\cite{shan2010survey}). As a result, the use of simulators is currently limited to what-if analysis. Typically filed data such as physical measurement of a simulated object of survey \cite{auld2012internet} can be used to calibrate parameters of a simulator. \cite{baker2020analyzing} provide an overview of Bayesian techniques for calibration and analysis of simulators.


\subsection{Bayesian Analysis of Computer Code Output (BACCO)}
Increases in computing power have generated an increasing desire to quantify uncertainty in statistical models. Even more so as these models become embedded into decision making. Current methods of sensitivity and uncertainty analysis usually require thousands of model runs, tracking changes of inputs and subsequent results, to form a distribution of outcomes. This sensitivity analysis allows model developers and decision makers to understand the relationships between inputs and outputs and risks and uncertainty surrounding a model. Normally, developers perform sensitivity and uncertainty analysis by changing the inputs and running the model, repeating this process potentially thousands of times. If the model is sufficiently complex, this process can become time-intensive. Though even a model that takes 2 seconds to run may require up to 6 hours to compute 10,000 runs.

To reduce the time it takes to perform necessary sensitivity and uncertainty analysis, Anthony O'Hagan developed a tutorial for using Bayesian Analysis of Computer Code Output (BACCO). He writes that rather perform thousands of runs on a model, which can be costly and time consuming, developers should choose to create an emulator, or a statistical approximation of the model (or the simulator). If this emulator is close enough to the actual model, it can be used to perform sensitivity and uncertainty analysis.

The process of building an emulator is relatively straight forward. O'Hagan makes use of a Gaussian process (GP). The GP uses a relatively small amount of simulator runs to calculate an estimate of the model. In the case of a single-input, single-output model, the GP will take the inputs and outputs used in the simulator runs and estimate the relationship using a multivariate normal distribution. The GP will produce exact results for the $X$s and $Y$s provided, but for all other combinations, it will specify a distribution, which is often visualized using 95\% confidence bounds. These bounds will increase as the unknown combinations move further from known combinations. In the case of a single-input, single-output model, the bounds will be widest at the midpoint between two points. In order to decrease the uncertainty of the emulator, the developer should choose points to minimize these bounds as much as possible, without performing too many simulator runs. This is usually not an issue for low-dimensional models. However, the more complex a model becomes, with more inputs and more outputs, it can be difficult to find good inputs such that the simulator runs will minimize emulator uncertainty. A plot of this process estimating a sinusoidal function is shown in the figure below.

\begin{center}
\includegraphics[scale = 0.75]{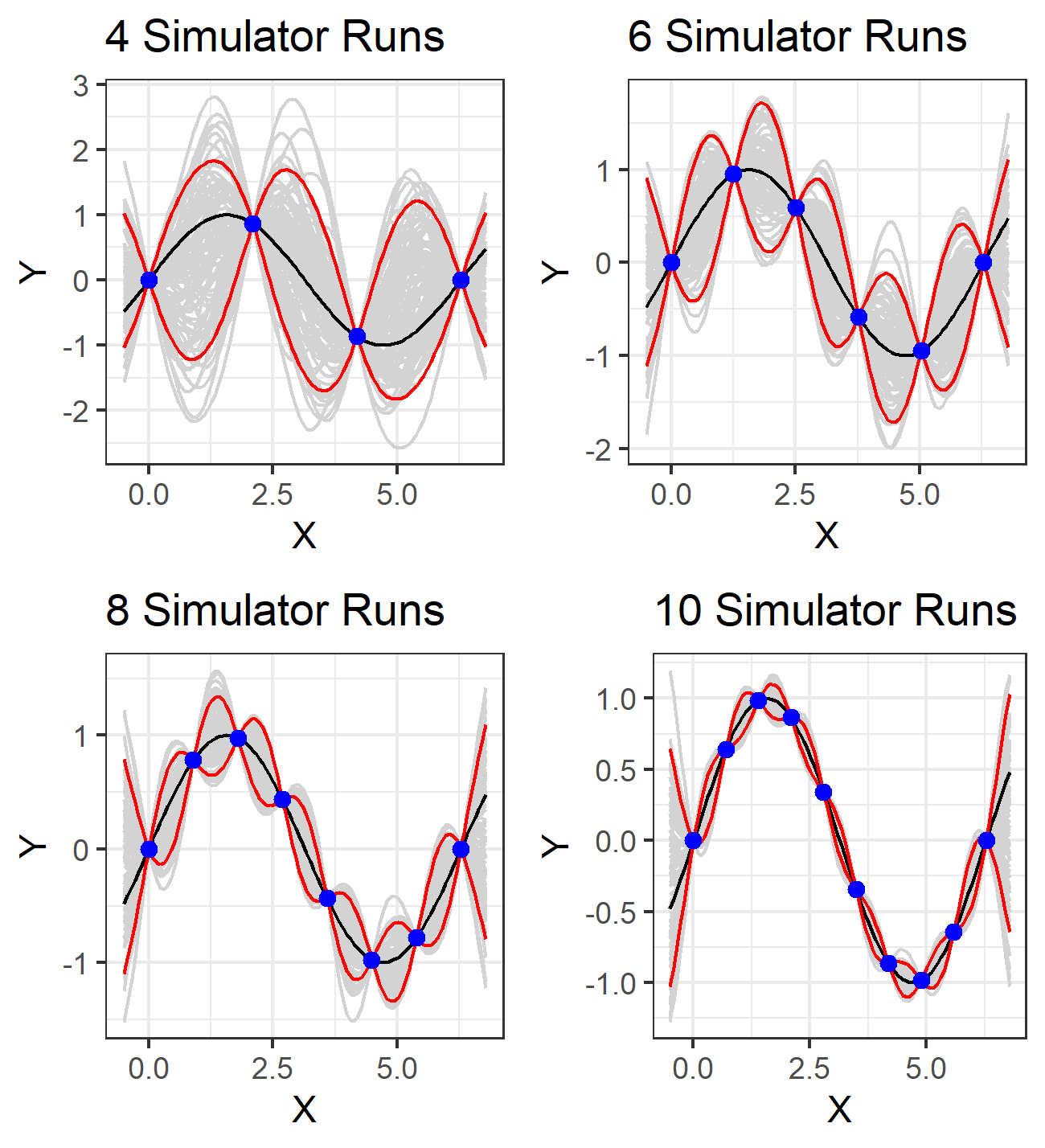}
\end{center}

Another advantage of using GPs is that they are inherently smooth. This is especially useful when trying to find optimality of the simulator. However, a downside to this smoothness is that the emulator has a "locally poor representation" (O'Hagan, 2004) of the simulator. This is a current area of research for BACCO.

To perform uncertainty analysis, we aim to find a distribution of the outputs. Often this is done using Monte Carlo methods on the simulator. O'Hagan writes "The simplest way to use the emulator is to treat the emulator's mean function $\hat{f}(.)$ as if it were the simulator $f(.)$, and just to apply the conventional methods to $\hat{f}(.)$." (O'Hagan, 2004) So we can still sample inputs from the input's distribution, but then we perform uncertainty analysis on the emulator outputs rather than the simulator's. For example, if we were trying to find the mean of the output's distribution, we would take the mean of the emulator's mean. However, Monte Carlo methods are not needed as we can derive the distribution of the outputs exactly.

BACCO methods are also very useful for model calibration, or "the process of using observational data to learn about uncertain model inputs" (\cite{ohagan_bayesian_2006})(O'Hagan 2004). Standard methods of calibration usually include iterating on inputs until the model output matches the new observational data. Once a close-enough match is found, these inputs are regarded as the true inputs. This methodology does not include any quantification of uncertainty. The Bayesian method of calibration explicitly defines uncertainty in the inputs. Rather than update the inputs directly, the Bayesian method of calibration updates the model parameter distributions to match the observed data. This will minimize uncertainties for inputs and gives output distributions where the observed data is most likely.

\subsection{Calibration of agent-based models for forecasting emerging infectious diseases}
It is difficult to calibrate the settings of agent-based models that require a large amount of computer processing time. Simulating the spread of infectious disease is one application of agent based models, but since calibration is time consuming it is important to create calibration techniques when there is not an outbreak, so-called "peacetime", so that during an outbreak (or "wartime"), the model can be used to make accurate predictions
    
As part of the Ebola forecasting challenge, an agent based model of the 2014 Ebola outbreak was calibrated using Bayesian calibration techniques and Simulation optimization. Simulation Optimization treats calibration like an optimization problem and looks to minimize the loss function in parameter space. In Bayesian calibration, sampled simulations runs are used to create the prior distribution. Simulation optimization is faster but has two primary drawbacks: The search may converge to a local minimum of the loss function, and that forecasts may only capture the stochasticity of the simulation and not parameter uncertainty. To account for those drawbacks, Bayesian calibration was also used

\cite{VenkatramananSrinivasan2017UDAm} proposed Nelder-Mead direct search method for simulation optimization. For the optimization approach it is crucial to choose the correct loss function that measures the error between simulated results and the observed Ebola outcomes. It was found that using the incremental Ebola observations with a weighted-L1 norm and reasonably high decay is what performed best during calibration.

The procedure used for Bayesian calibration is as follows: first in order to sample across the entire parameter space $\Theta$ a Latin hyper-cube sample of size 100 is taken over it. Simulations are carried out (replications 100 times) to create a digital library. For the parameter vector $\theta$, the prior distribution is assumed to be uniform over a specified rectangle. A response surface is fit to the digital library producing a mapping from the parameters to the simulated curve. Using this response curve, the likelihood of the observed simulation curve ($Y_i$) can be evaluated as a function of the model parameter setting. The data are assumed to be normal with a standard deviation  of 10\% and independent over each time period. The log likelihood can be expressed as:
\begin{equation}
    L(\theta|Y) = const - \frac{1}{2}\sum_{i=1}^{T}\frac{|X_i(\theta)-Y_i|^2}{0.1*X_i(\theta)}
\end{equation}
To be able to use convert the obtained posterior into a collection of Ebola results to be used in forecasting, two techniques were tested. 1) Using the posterior distribution to re-weight the Digital Library or 2) sample parameter configurations from the posterior and run fresh simulations

\subsection{Calibrating a Stochastic, Agent-Based Model Using Quantile-Based Emulation}
Agent-based models can be computationally demanding, and purely sampling based (Monte Carlo) approaches for sensitivity analysis  or inference make direct and repeated use of model output. In cases where a model run is too resource intensive, emulation is an effective approach. Emulation (also called surrogates or meta-modeling) is modeling the input output response of a model. One strategy is to model output distribution as normal, conditional on the mean and variance, and use a Gaussian process to model the mean and variance of the inputs. We will look at extending a quantile kreiging emulator to handle functional, stochastic model output.  

In another attempt to calibrate a model for the Ebola challenge, an Agent-based model from which only a limited numbers of simulation may be carried out was used by \cite{FadikarArindam2018CaSA}. This computer simulation uses a standard setup in Bayes computer model calibration where a group of ABM runs are carried out at input settings {$\boldsymbol\theta_1^*,...,{\boldsymbol\theta_m^*}$}. This group of runs is combined with observations $y$ from the actual pandemic using a Bayesian formulation discussed in the previous section. One key component of this formulation is modeling the simulation output $\eta(\boldsymbol\theta)$ using parameter settings $\theta$ that are not part of the group of ABM runs. To do this a Gaussian process (GP) is used

The input parameter for the Ebola challenge ABM has five dimensions and is represented  $\boldsymbol\theta = (\theta_1,...,\theta_5)$, and a collection of $m = 100$ of those settings is generated using a Latin Hyper-cube. For each setting the simulation is run 100 times. The model output is taken as the weekly number of of infected individuals and 57 logged counts per run. The varied output makes modeling the output as $N(\mu(\boldsymbol\theta),\sigma(\boldsymbol\theta))$ not work, so a quantile kreiging approach is used. GPMSA is the chosen software for the surrogate statistical modeling.

The basic implementation in GPMSA models the observation vector y as an additive combination of computational model $\eta(\boldsymbol\theta,\alpha)$ at the best setting for the parameters, a systematic model discrepency term $\delta$, and an observation error $\epsilon$ 
\begin{equation}
y = \eta(\boldsymbol\theta,\alpha) + \delta +\epsilon
\end{equation}

A Gaussian Process emulator is specified for the ABM output so that inference regarding model output at untried settings $\eta(\boldsymbol\theta,\alpha)$ can be accounted for in the analysis. In order to account for the multivariate nature of the model output, along with the observations, the basis representation is specified for $\eta()$
\begin{equation}
    \eta(\boldsymbol\theta,\alpha) = \phi_0 + \sum_{k=1}^{p_\eta}\phi_kw_k(\boldsymbol\theta,\alpha) + \epsilon_{w0}
\end{equation}
$p_\eta$ is the number of basis functions (for this application $p_\eta = 5$) and $\epsilon_{w0}$ is there to account for the error due to a limited number of basis functions. The basis functions are eigen vectors computed from the group of simulation run outputs, where $\psi$ is the mean of the curve produced by the output and $w_k$ are the Gaussian processes. 
Constructing the likelihood function is not straight forward due to the nature of the epidemic curves produced in the Ebola simulations and the observed Ebola epidemic curves. The log likelihood is modeled as
\begin{equation}
    \ell(\boldsymbol\theta,\alpha,\lambda_y;y,\eta(\cdot),\delta,\Sigma_y)= \frac{n_y}{2}\log\lambda_y-\frac{1}{2}(y-\eta(\boldsymbol\theta,\alpha)-\delta)^T\lambda_y{\Sigma_y}^{-1}(y-\eta(\boldsymbol\theta,\alpha)-\delta)
\end{equation}
The log likelihood form the ABM settings and the precision scaling term $\lambda_y$
 depends on the GP emulator $\eta(\cdot)$. This likelihood term for the observations is augmented with a likelihood term for the simulated epidemic curves.

\subsection{Uncertainty Quantification for Computer Models With Spatial Output Using
Calibration-Optimal Bases}

The calibration of simulations using Uncertainty Qualification (UQ) methods is a rich area of statistical development. When applying these techniques to simulators with spatial output, it is standard to use principal component decomposition to reduce dimension and allow Gaussian Process emulators (as discussed in the previous section). \cite{SalterJamesM2019UQfC} introduce "terminal cases" in which a model cannot reproduce observations to within model discrepancy, and presents a simple test to allow people utilizing simulations to establish whether their experiment will result in a terminal case. This section also looks at the optimal rotation algorithm for doing this

Two statistical methodologies this paper focuses on are Bayesian calibration and history matching with iterative refocusing. We will briefly discuss history matching and focus more on Bayesian calibration. Viewing a simulation a function with vector outputs $f(\boldsymbol\theta,\mathbf x)$ and inputs $\boldsymbol\theta$, and $\mathbf y (\mathbf x)$ is the simulated physical system, with $\mathbf z$ being observations of $\mathbf y$, both methodologies begin with the same assumption : that there exists a best input setting $\boldsymbol\theta^*$ such that
\begin{equation}
    \mathbf y = f(\boldsymbol\theta^*)+\boldsymbol\eta,    \mathbf z = \mathbf y + \mathbf e
\end{equation}
where $\mathbf e$ are observation errors and $\boldsymbol\eta$ represents model discrepency. Both methodologies require a Gaussian process emulator

The method of history matching and iterative refocusing allows for validation of the model as part of the calibration exercise. It alters the problem of calibration to a problem of trying to rule out regions of the parameter space $\Theta$ that do not contain the optimal inputs. An invalid model would have all of $\Theta$ ruled out. History Matching and iterative refocusing defines an implausibility function

\begin{equation}
    \mathcal{I}(\boldsymbol\theta) = (\mathbf z - E[f(\boldsymbol\theta)])^T (\text{var}(\mathbf z - E[f(\boldsymbol\theta)]))^{-1}(\mathbf z - E[f(\boldsymbol\theta)])
\end{equation}

the expected values and variances of $f(\boldsymbol\theta)$ are derived from a Gaussian process emulator. and are conditioned on the model runs $\mathbf F =( f(\boldsymbol\theta_1),...,f(\boldsymbol\theta_n))$ .If $\mathcal{I}(\boldsymbol\theta)$ exceeds some threshold T, those parameters are ruled out. Everything else is a member of the subspace $\Theta'$ and Not Yet Ruled Out (NYRO)

From a history matching persepective, the terminal case occurs when the model is too far from the observations at every point in $\Theta$. Within a probabilistic framework, the terminal case implies a prior-data conflict so that the discrepancy variance assessment has been misspecified, or the expert is wrong. The terminal case can occur even in models that can reproduce observations exactly

For spatial fields, the most common approach to emulation and calibration involves projecting the model output onto a low dimensional basis $\boldsymbol\Gamma$, and emulating the coefficients, so that fewer emulators are required. This was discussed in the previous section briefly when discussing the basis representation. Writing the model output as a vector of length $\ell$ so that $\mathbf F$ has dimension $\ell\times n$, the singular value decomposition is used to give $n$ eigen vectors that can be used as basis vectors. The majority of variability in $\mathbf F$ is in the first few eigen vectors, so the basis is truncated after $q$ vectors giving $\boldsymbol\Theta_q$ a dimension of $\ell \times q$. For emulation to occur, the model runs are centered by subtracting their mean from each column of $\mathbf F$ giving a centered group of runs $\mathbf F_\mu$  which is then projected on to $\boldsymbol\Gamma_q$ from which the emulators can be constructed. Given thees emulators, calibration can either be performed using the entire $\ell$-dimensional output, or on its $q$-dimensional basis representation

For calibration there are two main requirements for a basis $\mathbf B$

\begin{itemize}
    \item Being able to represent $\mathbf z$ with $\mathbf B$ 
    \item Retaining enough signal in the chosen subspace to enable accurate emulators
\end{itemize}
Natural methods satisfying the first goal is to minimize error given when the observations are reconstructed using $\mathbf B$. We define reconstruction error as
\begin{equation}
    \mathcal{R}_W(\mathbf B, \mathbf z) = \|\mathbf z - \mathbf B (\mathbf B^T \mathbf W^{-1}B)^{-1}\mathbf B^T\mathbf W^{-1}\mathbf z \|_\mathbf W
\end{equation}
$\|\mathbf v\|_\mathbf W=\mathbf v^T\mathbf W^{-1}\mathbf v$ is the norm vector of $\mathbf v$ and $\mathbf W$ is an $\ell \times \ell$ weight matrix. By setting $\mathbf W = \boldsymbol\Sigma_e + \boldsymbol\Sigma_\eta$, $\mathcal{R}_W(\mathbf B, \mathbf z)$ is analogous to $\mathcal{I}(\boldsymbol\theta)$. For the history matching threshold T, if $\mathcal{R}_W(\mathbf B, \mathbf z) > T$, then we are on the terminal case pn $\mathbf B$. If $\mathcal{R}_W(\mathbf B, \mathbf z) > T$ for some $\{\mathbf B, \mathbf W \}$, then $\mathbf W$ is misspecified

To find the optimal basis for performing calibration and testing for the terminal case, the following algorithm can be used to rotate the basis $\mathbf B$ to the optimal basis
\begin{enumerate}
    \item If $\mathcal{R}_W(\mathbf B, \mathbf z) > T$ , stop and examine the specified $\mathbf W$ or increase the amount of runs in $\mathbf F_{\boldsymbol\mu}$, else set $k=1$
    \item let $\boldsymbol\Gamma_k^* = (\boldsymbol\gamma_1^*,...,\boldsymbol\gamma_{k-1}^*,\mathbf B\boldsymbol\lambda_k)$ and set 
    
    $\boldsymbol\lambda_k^* = \text{arg min}_{\boldsymbol\lambda_k }\mathcal{R}_W(\boldsymbol\Gamma_k^*, \mathbf z)$
    
    such that $\mathcal{V}_k(\boldsymbol\Gamma_k^*,\mathbf F_{\boldsymbol\mu} ) \geq v_k$. Define the new normalized vector as 
    
    $\boldsymbol\gamma_{k
    }^* = \frac{\mathbf B\boldsymbol\lambda_k^*}{\sqrt{\|\mathbf B\boldsymbol\lambda_k^*\|_\mathbf W}}$
    
    and set $\boldsymbol\Gamma_k^* = (\boldsymbol\gamma_1^*,...,\boldsymbol\gamma_{k-1}^*,\boldsymbol\gamma_k^*)$
    \item Find the residual basis given $\boldsymbol\Gamma_k^*,\mathbf B_\epsilon^k$, and form the orthogonal rank n basis
    
    $\boldsymbol\Gamma^* = (\boldsymbol\Gamma_k^*,[\mathbf B_\epsilon^k]_{n-k})$
    
    \item Define $q\geq k$ as the minimum value satisfying $\mathcal{V}_k(\boldsymbol\Gamma_q^*,\mathbf F_{\boldsymbol\mu} ) \geq v_{\text{tot}}$, where $\boldsymbol\Gamma_q^*$ represents the first q columns of $\boldsymbol\Gamma^*$. If $\mathcal{R}_W(\boldsymbol\Gamma_q^*, \mathbf z) < T$, then stop, and return $\boldsymbol\Gamma_q^*$ as the truncated basis for calibration. Else set $k=k +1$ and $\mathbf B = [\mathbf B_\epsilon^k]_{n-k}$, and return to step 2 (\cite{SalterJamesM2019UQfC})
 \end{enumerate}

\subsection{Combining field data and computer simulations for calibration and prediction}
\cite{HIGDONDave2005Cfda} formulate a general statistical approach for combing scant field observations with simulator runs to calibrate parameters in the simulator and to character uncertainty in simulator-based predictions. This approaches explicitly models uncertainty in model inputs, for limited numbers of simulation runs, and for discrepancy between the simulation and the actual physical system. This approach uses standard statistical models including Gaussian process models for convenience, flexibility and generality 

Simulator output of input vector $(x,t)$ is denoted as $\eta(x,t)$. At various settings for $x$, observations are made, denoted as 
\[
    y(x_i)=\zeta(x_i) + \epsilon(x_i), i=1,...,n
\]
$\epsilon$ represents observation error. For a system with multivariate observations, each y can still be univariate and modeled statistically with $\eta(x,\theta)$, $\theta $ is the true calibration value in
\[
    y(x_i)=\eta(x_i,\theta) + \delta(x_i)+ \epsilon(x_i), i=1,...,n
\]
where $\delta$ accounts for model discrepancy, and $\theta$ is the "true" or "best" value for calibration

For a situation with unlmitied simulation runs, the model $y(x_i)=\eta(x_i,\theta) + \epsilon_i, i=1,...,n$ is appropriate and the sampling model for y is expressed as 
\[
L(y|\eta(\theta)) \propto \exp \Big\{ -\frac{1}{2}(y-\eta(\theta))^T\Sigma_y^{-1}(y-\eta(\theta))\Big\}
\]
To complete the Bayesian formulation, a prior distribution $\pi(\theta)$ is specified. The resulting posterior distribution for the parameters $\theta$ is thus
\[
    \pi(\theta|y) \propto L(y|\eta(\theta)) \times \pi(\theta) \propto \exp \Big\{ -\frac{1}{2}(y-\eta(\theta))^T\Sigma_y^{-1}(y-\eta(\theta)) - \frac{1}{2(.25^2)}(\theta-.5)^2\Big\}
\]
Next, a fairly general purpose is to generate a sequence of realizes of $\theta$, and this can be carried out using the MCMC method, and a simple implementation uses the Metropolis algorithm
\begin{enumerate}
    \item Initialize $\theta^1$ at some value
    \item Given the current realization is $\theta^t$, generate $\theta^*$from a symmetric distribution
    \item Compute metropolis acceptance probability
    
    \[\alpha = \min \Bigg\{1, {\frac{\pi(\theta^*|y)}{\pi(\theta^t|y)}}\Bigg\}\]
    
    \item Set
    \[
        \theta^{t+1} = \left\{\begin{array}{cc}
        \theta^* & \text{with probability } \alpha \\ 
        \theta^t & \text{with probability } 1 - \alpha
        \end{array} \right.
    \]
    \item Iterate over steps 2-4
    \end{enumerate}
    
When the model complexity makes unlimited runs impossible, the above approach that requires large numbers of runs, will not work. As discussed in previous section, a Gaussian process emulator can be used. A mean functions and a covariance are required to fully specify a GP prior model. The covariance can be specified as
\[
    \text{Cov}(x,t),(x^{'},t^{'})) = \frac{1}{\lambda_\eta}\exp\Bigg\{
    -\sum_{k=1}^p \beta_k^\eta|x_{ik}-x_{ik}^{'}|^\alpha -
    \sum_{k^{'}=1}^\ell \beta_{p+k^{'}}^\eta|t_{ik^{'}}-t_{ik^{'}}^{'}|^\alpha
    \Bigg\}
\]
Then the likelihood  is
\[
    L(z|\theta,\mu,\lambda_\eta, \beta^\eta, \Sigma_y) \propto
    |\Sigma_z|^{-\frac{1}{2}}\exp\Bigg\{-\frac{1}{2}(z-\mu1_{n+m})^T\Sigma_z^{-1}(z-\mu1_{n+m})\Bigg\}
\]
where $1_{n+m}$ is the n + m vector of 1s. After using a Gaussian process the posterior distribution is thus
\[
    \pi(\theta,\mu,\lambda_\eta,\beta^\eta|z) \propto L(z|\theta,\mu,\lambda_\eta, \beta^\eta, \Sigma_y)\pi(\theta)\pi(\mu)\pi(\lambda_\eta)\pi(\beta^\eta)
\]

Finally to account for model discrepancy, that is cases where field observations are inconsistent with simulations no matter the input. to account for that discrepency another term is added to $y$, $\delta(x_i)$ so that y becomes
\[
 y(x_i)=\eta(x_i,\theta) + \delta(x_i)+ \epsilon(x_i), i=1,...,n
\]
And in this case the new covariance for the Gaussian Process is
\[
\text{Cov}(x,x\prime) = \frac{1}{\lambda_\delta}\exp \Bigg\{-\sum_{k=1}^p \beta_k^\delta |x_{ik}-x_{ik}\prime|^{\alpha_\delta}\Bigg\}
\]
And the prior is now
\[
\pi(\lambda_\delta) \propto \lambda_\delta^{\alpha_{\delta}-1}\exp\{-b_\delta \lambda_\delta\}, \lambda_\delta>0
\] 
and
\[
\pi(\beta^\delta) \propto \prod_{k=1}^p (1-e^{-\beta_k^\delta})^{.6}e^{-\beta_k^\delta}, \beta_k^\delta > 0 
\]
The likelihood and posterior are the same as previously stated for dealing with limited simulation runs
\subsection{Model Evaluation and Spatial Interpolation by Bayesian Combination of Observations with Outputs from Numerical Models}
\cite{FuentesMontserrat2005MEaS} consider models  that also have spatial outputs and need to be calibrated. Spatial output models can be used to model air pollution. One of the problems is that the observations may not be the actual ground truth as there can be sensor error, so we will look at a simple Bayesian model for the unobserved ground truth. This will allow for improved spatial prediction via a posterior distribution for ground truth.

When measuring pollution, we cannot consider those measurements to be ground truth due to measurement error. Instead, it is assumed there is an unobserved field Z(s) and the measurement is $\hat{Z}(s)= Z(s) + e(s)$ where $e(s)$ is that measurement error. The true underlying process Z is assumed to be $Z(s) = \mu s(s) + \epsilon(s)$ where Z(s) has a spatial trend $\mu(s)$ that is polynomial. For spatial prediction, we simulate values of Z from is posterior predictive distribution a $P(Z|\hat{Z},\Tilde{Z})$

The algorithm for estimation and prediction of a spatial model uses a Gibbs sampler. We alternate between the parameters that measure lack of stationarity, and we obtain the conditional posterior distribution from those parameters. Given the values of Z that are updated in the last stage, the posterior distribution will be completely specified once we define the prior because $(Z|\beta, \theta)$ is Gaussian

\subsection{Numerical Example of Surrogates}
We will now look at a numerical example of surrogates in action. Derived from examples by \cite{gramacy2020surrogates}, we will look at using surrogates when there is a black box function. To do this we will consider a function $f(x)$. The function we will consider is not in reality a black box, nor is it computationally expensive, but we need a function to evaluate and so we use the following for $f(x)$
\begin{verbatim}
    f <- function(X)
    {
      if(is.null(nrow(X))) X <- matrix(X, nrow=1)
      m <- 8.6928
      s <- 2.4269
      x1 <- 4*X[,1] - 2
      x2 <- 4*X[,2] - 2
      a <- 1 + (x1 + x2 + 1)^2 * 
        (19 - 14*x1 + 3*x1^2 - 14*x2 + 6*x1*x2 + 3*x2^2)
      b <- 30 + (2*x1 - 3*x2)^2 * 
        (18 - 32*x1 + 12*x1^2 + 48*x2 - 36*x1*x2 + 27*x2^2)
      f <- log(a*b)
      f <- (f - m)/s
      return(f)
    }
\end{verbatim}
To illustrate the use of surrogates we will start with a Latin hyper-cube sample so that we have samples across the sample space

\begin{verbatim}[language=R]
ninit <- 12
X <- randomLHS(ninit, 2)
y <- f(X)
\end{verbatim}

next we fit a GP emulator 

\begin{lstlisting}[language=R]
da <- darg(list(mle=TRUE, max=0.5), randomLHS(1000, 2))
gpi <- newGPsep(X, y, d=da$start, g=1e-6, dK=TRUE)
mleGPsep(gpi, param="d", tmin=da$min, tmax=da$max, ab=da$ab)$msg
\end{lstlisting}

\begin{verbatim}[language=R]
obj.mean <- function(x, gpi) 
  predGPsep(gpi, matrix(x, nrow=1), lite=TRUE)$mean
m <- which.min(y)
opt <- optim(X[m,], obj.mean, lower=0, upper=1, method="L-BFGS-B", gpi=gpi)
opt$par
\end{verbatim}
And we use that to come up with a new point guess to optimize the computationally costly black box function. Here is a visualization
\begin{center}
  \includegraphics[scale=.5]{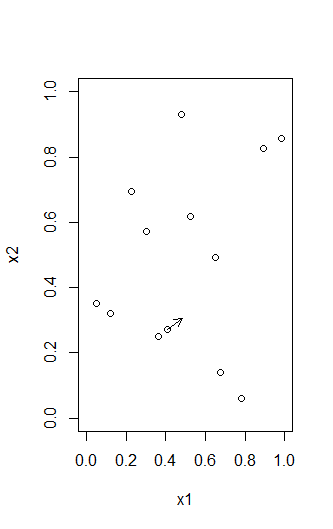} 
\end{center}

The arrow originates from the location whose y value is lowest based on the initial evaluations and it is pointing at the next point to try. We now  find $f$ at the the value of opt\$par and reiterate to get a new optimization guess
\begin{verbatim}[language=R]
ynew <- f(opt$par)
updateGPsep(gpi, matrix(opt$par, nrow=1), ynew)
mle <- mleGPsep(gpi, param="d", tmin=da$min, tmax=da$max, ab=da$ab)
X <- rbind(X, opt$par)
y <- c(y, ynew)
m <- which.min(y)
opt <- optim(X[m,], obj.mean, lower=0, upper=1, method="L-BFGS-B", gpi=gpi)
opt$par
\end{verbatim}

We once again plot the new optimization guess
\begin{center}
  \includegraphics[scale=.5]{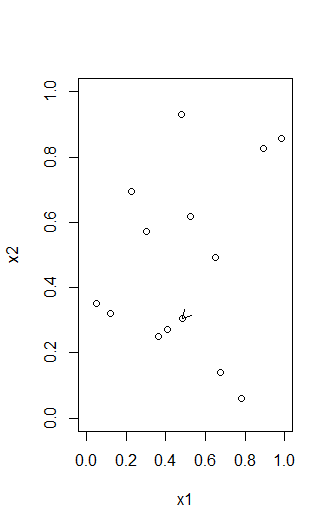}  
\end{center}

Now to fully explore this example we use an R function to automate the process and produce more optima
\begin{verbatim}[language=R]
optim.surr <- function(f, m, ninit, end, tol=1e-4)
 {
  ## initialization
  X <- randomLHS(ninit, m)
  y <- f(X)
  da <- darg(list(mle=TRUE, max=0.5), randomLHS(1000, m))
  gpi <- newGPsep(X, y, d=da$start, g=1e-6, dK=TRUE)
  mleGPsep(gpi, param="d", tmin=da$min, tmax=da$max, ab=da$ab)

  ## optimization loop
  for(i in (ninit+1):end) {
    m <- which.min(y)
    opt <- optim(X[m,], obj.mean, lower=0, upper=1, 
      method="L-BFGS-B", gpi=gpi)
    ynew <- f(opt$par)
    if(abs(ynew - y[length(y)]) < tol) break
    updateGPsep(gpi, matrix(opt$par, nrow=1), ynew)
    mleGPsep(gpi, param="d", tmin=da$min, tmax=da$max, ab=da$ab)
    X <- rbind(X, opt$par)
    y <- c(y, ynew)
  }

  ## clean up and return
  deleteGPsep(gpi)
  return(list(X=X, y=y))
 }
 
 
\end{verbatim}
Now lets run this 1000 timees, but we will see that even with 1000 Monte Carlo repetitions convergence is often found by 80 or even 60 repetitions
\begin{verbatim}
reps <- 1000
end <- 80
prog <- matrix(NA, nrow=reps, ncol=end)
for(r in 1:reps) {
  os <- optim.surr(f, 2, ninit, end)
  prog[r,] <- bov(os$y, end)
}
\end{verbatim}
And here we can see the 1000 trajectories stored in the prog variable
\begin{center}
    \includegraphics[scale=.5]{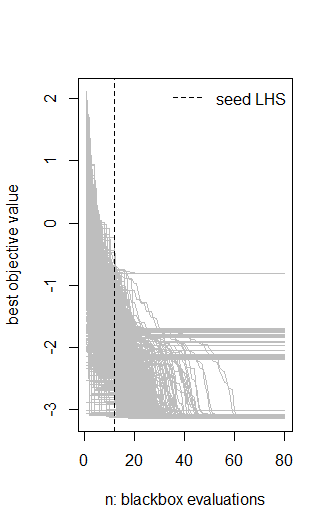}
\end{center}
That is a quick numerical example of optimization done with a Gaussian Process surrogate. There are better versions with some more refinements but to give a brief good example this is what we present \cite{gramacy2020surrogates}

\section{Applications}

\subsection{Causal Learning}
Data modeling using descriptive, prescriptive, and predictive technologies proves a means to understand, visualize, recognize correlated variables and predict future trends. But these models fails to uncover the direct or indirect causal and effect relationships among variables. This identification of causality can be performed by causal discovery models. Causal Learning is an emerging technology used in many fields to infer cause and effect relationships among the interacting components of the main system. \cite{ARORA2019439} utilized causal learning method to uncover the cause of certain disease. \cite{li2013causality} studied how the factor Foreign Direct Investment (FDI) influenced the economy of different countries including their social, political and other institutional factors, using a Directed Acyclic Graph (DAG) approach. 

Causal Discovery works on the principle of diagrammatically representing variables as nodes, connected through arrows that represent the causal relationship amongst them. Direction of these arrows (edges) gives more insight on whether the variable is caused or effected. The underlying causality is determined the probability distribution of the nodes (variables) conditional on the adjacent nodes. 

\subsubsection{Directed Acyclic Graphs}

A sample Directed Acyclic Graph representation of three variables $X_1$, $X_2$, and $X_3$ is shown in the Figure 2. 

\begin{figure}[htbp] 
	\centering
	\includegraphics[scale=0.2]{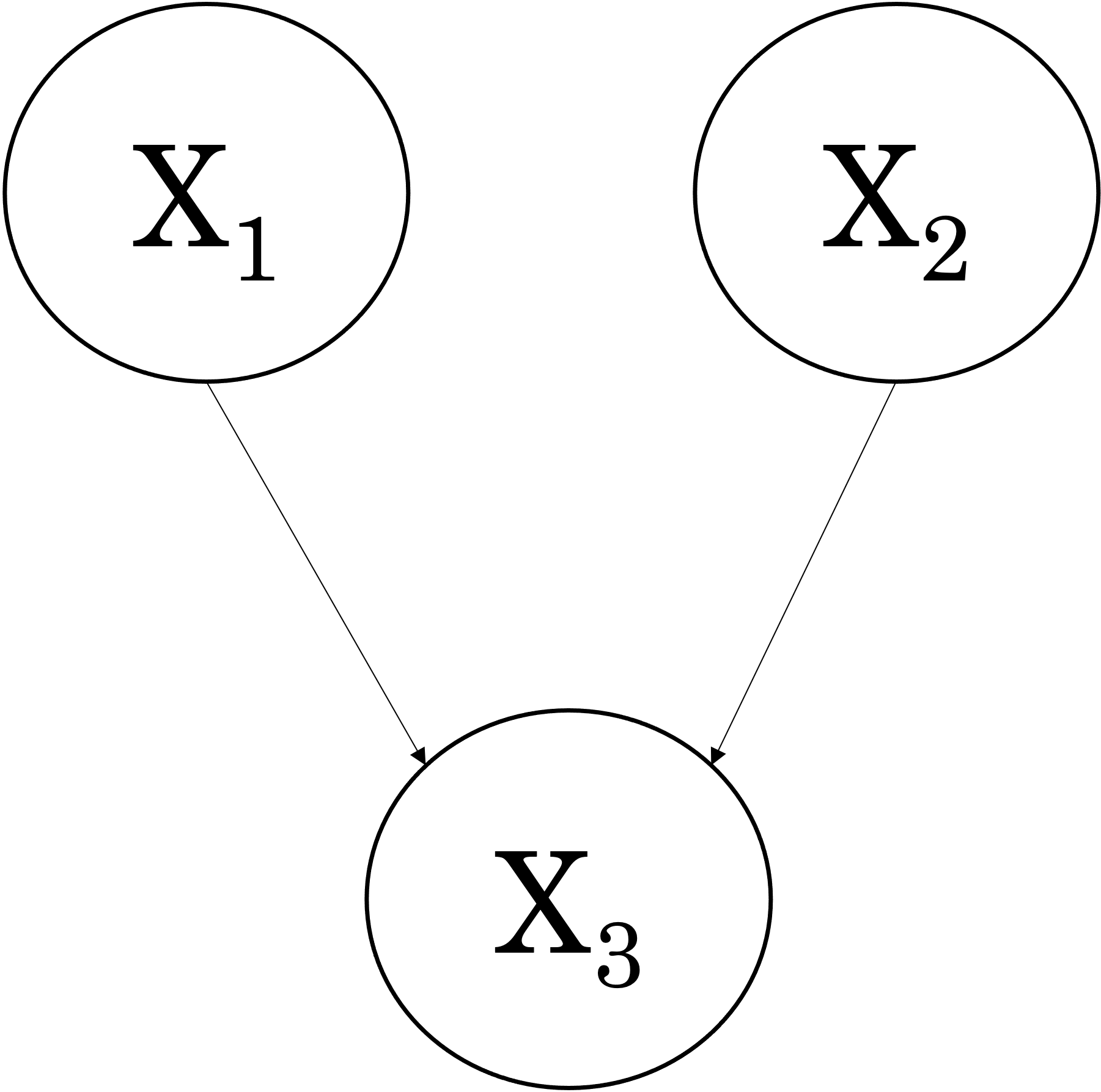} 
	\caption{Example Causal Diagram} 
	\label{fig.causal_dia} 
\end{figure}

Here, the nodes $X_1$ and $X_2$ are directed toward the node $X_3$, and thus means that the variables $X_1$ and $X_2$ together cause the effect on the node $X_3$. A DAG represented as a Graph $\mathbb{G}=(N,E)$ with a Node set $N = {N_1, N_2, .... N_i}$ and Edge set $E = {E_1, E_2, ....E_i}$ and a probability distribution $P$ and works under the principle of Causal Markov Condition. The condition states that the G and P satisfy the condition only if every node $N_i$ in the set N is independent of its non-parent nodes given its Parents. Numerous techniques are in use to render a DAG. Each of them can be under the umbrella of any one of the following - Constraint-based, Score-based and Hybrid causal learning methods. Constraint-based methods starts with an undirected graph of all variables in the dataset and direct the arrows (edges) based on testing the probability of the nodes given their parents. Similarly, Score-based methods also starts with a graph and orient the arrows based on the a score (such as log likelihood). Combining the best of both techniques are the hybrid causal learning methods. Many variants of DAGs have been developed in many fields. Bayesian Network is one such variant of DAG where BN is a network diagram representation that works on a conditional probability distribution of the nodes.

\subsection{Bayesian Networks}
Let set Y = {$Y_1$, $Y_2$, .... $Y_n$} and each of the $Y_i$ variable takes a finite value. Then the set of nodes and edges consisting of the variable set Y be represented as  Node set $N = {N_1, N_2, .... N_i}$ and Edge set $E = {E_1, E_2, ....E_i}$.

A Bayesian is a directed acyclic graph similar to the Figure 1, where each node refer to the respective variable in the set Y. And the flow of causal information is presented through an independence relationships between the variables and their ancestors (parents). In BN each node has a marginal as well as conditional probability distribution. The conditional probability given its parents is an essential part of Bayesian Network. 

The conditional probability of a node conditional on its parent node is given by

\begin{equation}
 P(Y_i\mid Parent(Y_i)) = P(Y_i|Pa_i)
 \end{equation},

where $Pa_i$ is the parent node of the node $Y_i$. 

In the regression-based models, the parameter of interest is nothing, but an objective number represents the estimate of interest by minimizing the error $\epsilon$. Whereas, in BN the parameter of interest is the measure of belief.They calculate the estimates by CPD of the variable given the values of other variables. For e.g., P($Y_1 \mid Y_2= C$), where C is a constant of certain value. In BN, the PD is joint, and all the variables are modeled into the estimation. A full joint distribution of variables given by Bayesian Network is

\begin{equation}
 P(Y_1,...X_n)) = \Pi P(Y_i\mid(Pa_i))
 \end{equation}

BN can also be modelled with prior data and this serves very well in situations when data is sparse. Similarly, the Bayesian Networks perform belief updating when new information is obtained using Bayes Theorem. Hence they are also called as Bayesian Belief Networks. This updated probability of the nodes (variables) $Y_i$ given the dataset $A$ is determined by the following relationship:

 \begin{equation}
p(Y_m\mid A) = \frac{p(A\mid Y_m)*P(Y_m)}{\sum \limits_{N=1}^n p(A\mid Y_n)*P(Y_n)}
\end{equation} 

where $P(Y_m\mid A)$ is the posterior probability of Y based on the evidence or data E; P($Y_m$) refers to the prior probability; P($A\mid Y_n$) denotes the conditional probability of A considering $Y_m$.

\subsection{Bayesian Networks for Risk Prediction}

Risk Prediction is an essential technique used to predict the risk of an event. Many regression based and machine learning based approaches have been used for risk prediction. \cite{ARORA2019439} utilized Bayesian Network for risk prediction. Regression based has an advantage of being flexible, but they are not capable of showing the causal relationships/structure. Whereas, Bayesian Network has shown ability to predict even individual level risk and even perform what if scenarios. It can also be applied in situations where required data is unavailable. On the contrary to linear regression or even neural networks, in BNs it is possible to visually see and understand how each node probability is determined and which are the causal/contributing variables (nodes).

\begin{figure}[htbp] 
	\centering
	\includegraphics[scale=0.47]{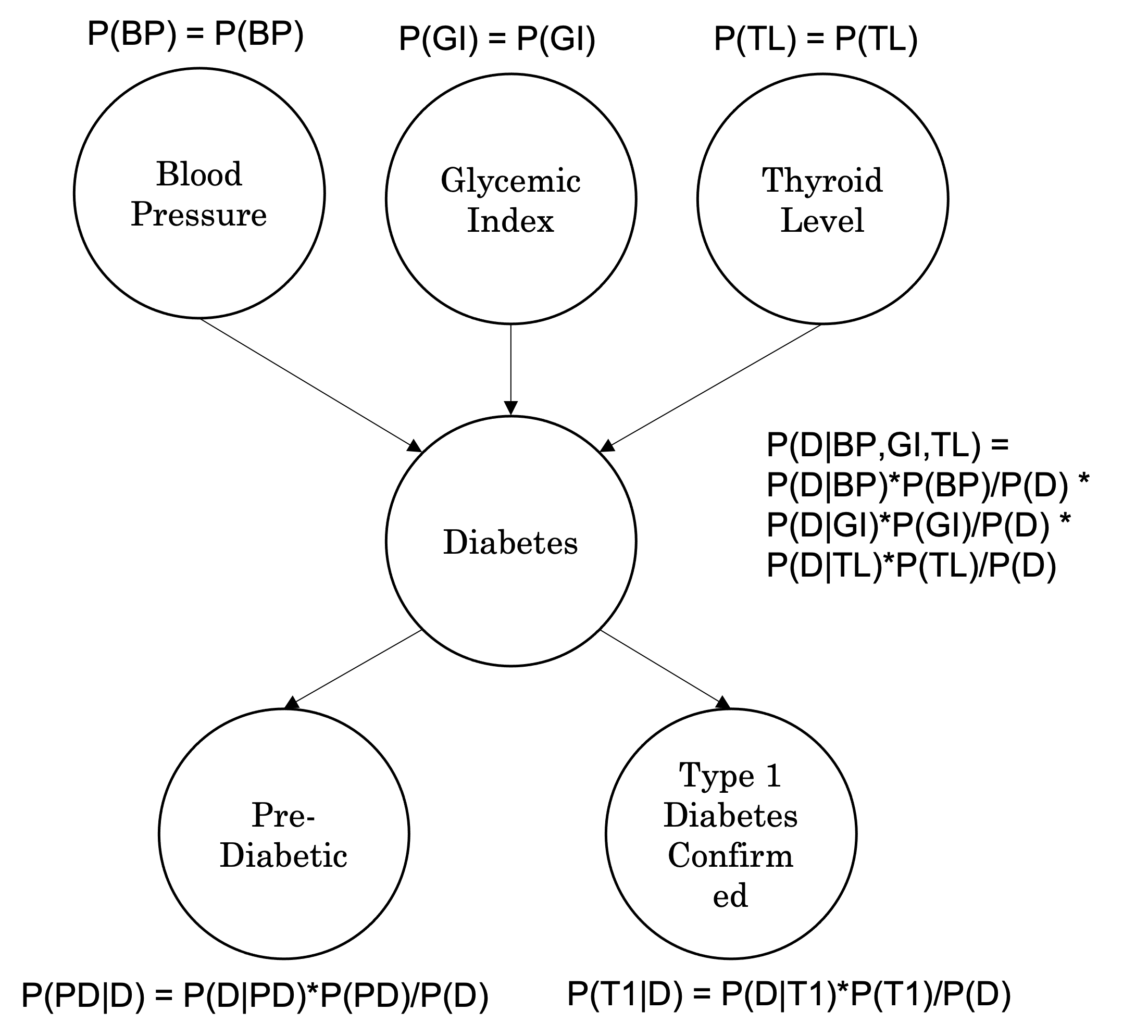} 
	\caption{Three Characteristics-Disease Outcome Causal Diagram} 
	\label{fig.disease} 
\end{figure}

In this study, \cite{ARORA2019439} utilized DAG representation to show how patient characteristics cause the outcome of disease in the test results. Figure 3, is a similar DAG representation of Three Patient characteristics namely - Blood Pressure, Glycemic Index and Thyroid Level influence the likelihood of the disease "Diabetes" outcome and in turn that affects the status of the test result - Pre-Diabetic or Type 1 Diabetes. 

The authors presented a scenario which predicts the risk of lung cancer and tuberculosis. Factors like visit to Asia, smoking, X-ray result etc. are included in the BN analysis. The advantage of BN is that it allows for modeling with or without prior information. When no prior information is available, non-informative priors are used for calculation of PD. BN provides a way to perform individual level what if analysis. It also helps decision makers to make informed decisions including the causal relationships. BNs are capable of handling both linear and non-linear relationships and also performs well with high dimensional data. However, as more variables are present in the dataset, the computational time of the BN also increases.

\subsection{Assessing areas of Vulnerability using Bayesian Belief Network}

Bayesian Belief Networks (BBN) are used in developing network diagrams with help of expert knowledge CPT (Conditional Probability Table) elicitation. \cite{ABEBE20181629} used this technique for developing a BBN and predicting Flood Vulnerability Index (FVI) in the city of Toronto. Pluvial Flooding has been ever increasing owing to the increased intensity and frequency of rains (because of climate change), loss of water storage space etc., in urban areas. These factors result in a city being vulnerable to flooding. Measuring this ``Vulnerability index'' would help city officials to plan for risk mitigation and adaptation. \cite{ABEBE20181629} have proposed a methodology to quantify uncertainty and causal links in the dataset. They then applied the methodology to a case study to diagnose variations in basement flooding in city of Toronto and also predicted Flood Vulnerability Index (FVI). Identifying various factors influencing the flooding were identified through subject matter experts and available literature. Data was then collected (22 parameters). Area of the city of Toronto was classified into smaller grid areas of 1 Sq Km and Grid specific variables are extracted including the no. of basement flooding. The new feature FVI is calculated from the no. of reported basement flooding. The authors used a tool called – Netica for BBN and and data was visualized using the GIS tool -  ArcGIS.

Sensitivity of a node can be quantified by many ways - variance reduction, mutual information, or variance of beliefs.\cite{ABEBE20181629} quantified the sensitivity of a node through variance reduction method. Sensitivity analysis (SA) was also carried out in the Netica software. SA to the node FVI was carried out to specifically see the order of influence of the other nodes on FVI. From the results, population density was found to be the most influential factor. FVI for future is predicted using the factors identified as major influencers. Authors validated the predicted FVI based on the values obtained from basement flood subsidy protection program (BFSPP). The R2 of the fitted model was 0.56. It was recommended that if missing parameter such as infrastructure condition were available R2 could be improved. Uncertainty of the prediction was quantified by probability. For e.g., for a grid 562, prediction was that it has more than 0.75 probability of having FVI of three or more (high or very high). The model reliability is affected by the data availability and quality

Owing to the lack of availability of data certain parameters like drainage infrastructure type, condition and capacity that were shown in BN DAG were not included in the final analysis by the authors. If these data were good and had better availability, then there would be better prediction of FVI and better causality analysis
This BBN can be applied to other numerous applications

\subsection{Consequence-based Bayesian Belief Network for buried infrastructure systems}

\cite{kabir2018consequence} used the BBN based consequence index model for assessing buried infrastructure. Buried infrastructures in a city has crucial roles to play in regard to proper functioning of the city. Mis-operation/repair of these structures will cause serious consequence to health and safety, environment, society, and economy and organization. \cite{kabir2018consequence} has developed a BBN based buried infrastructure consequence model to assess consequence index and prepare for maintenance of these structures
The authors utilized various available literature and expert knowledge to construct all contributing factors to the perform causal analysis
Consequence index is built by considering all other impact indexes such as health and safety impact (HSI), environmental impact (EVI), social impact (SCI), economical and organizational impact (EOI) 
First the features (10) from which these indexes can be extracted are constructed - Sewer diameter (SD) (mm), Sewer length (SL) (m), burial depth (BD) (m), Population affected (PA) (Number of person), etc.,. The causal relationship between the features and the indexes are constructed with help of expert knowledge and literature. The BBN-based consequence assessment model - causal structure was then developed using the help of Netica software. Here CPT is generated manually with help of SMEs and from the manually produced causal structure. CPT was generated through expert knowledge elicitation from 5 of the experts. CPT values kept changing until all 5 of the experts came to a consensus. The proposed model was validated through extreme-condition test and scenario analysis.

Two extreme scenarios were considered where Extreme 1 depicts all parent nodes in worst possible condition and Extreme 2 depicts all parent nodes in good/favorable conditions. When the proposed model was applied to this scenario, Extreme 1 and 2 Showed the CI to be 77.5\% and 60.1\% respectively. For the scenario analysis, the authors considered what-if analysis by changing the values of specific parameters like sewer length. These scenarios too resulted in the expected CI index values based on the framework. To identify the critical parameters and quantify model validity, sensitivity analysis is used. It quantify the model uncertainty by – presenting how subtle the (sensitive) information of the interested feature (o/p of the model) is when there were slight changes to all other input parameters. Variance Reduction method is used for this purpose. This method find the amount of variance of variable of interest (CI) reduced because of a varying feature in the study (Length, Diameter) etc.,. Thus, variance of the real value of CI, given an evidence N is calculated using the formula

\subsection{Bayesian Model Averaging}

Bayesian Model Averaging (BMA) is an ensemble/stacking technique to average the outputs of different models. This technique helps with improving the prediction of outcome. BMA uses the posterior probability distributions of the models and to averages across all the models considered to get a more efficient and informative output. It also helps in dealing with model uncertainty.

\cite{zou2021application} used BMA technique for analyzing freeway incident clearance time for emergency management. The dataset used in the study was the traffic incident observation data from the I-5 corridor between Boeing Access Road and the Seattle Central Business District were retrieved from the WITS. 2,584 valid incidents from 1 January to 31 December 2009 were selected from the WITS dataset including 15 categorical candidate explanatory variables. Find key factors that affect traffic incident clearance time

Different linear regression models were built for different predictor variable combinations to predict the response variable – incident clearance time. 20 different models with the highest PMP were finally selected by Occam's window approach. BMA is then applied to the chosen models. Posterior effect probabilities, posterior means, standard deviations are extracted for these models using BMA.The posterior effect probabilities of seven explanatory variables equal 100\%. Response time, traffic control, collision, multiple lane blocked, total closure, injury involved, and summer are the important factors affecting the duration of traffic incident clearance. The posterior effect probability in the BMA model helps to overcome the overstatement of the evidence for an effect. In certain models like Cox PH, the low p values for some variables indicated highly significant effect, whereas the BMA model suggested that the evidence for effect is not so strong.

\printbibliography
\end{document}